\documentclass[superscriptaddress, aps, pra, twocolumn, showkeys]{revtex4-2}

\usepackage[colorlinks=true, citecolor=blue, urlcolor=blue, linkcolor=magenta]{hyperref}
\usepackage{braket}
\usepackage{amsmath,amssymb,amsthm}
\usepackage{easybmat}
\usepackage[colorlinks=true,citecolor=blue,urlcolor=blue, linkcolor=magenta]{hyperref}
\usepackage[pdftex]{graphicx}
\usepackage{times,txfonts}
\usepackage{braket}
\usepackage{color}
\usepackage{xcolor}
\usepackage{natbib}
\usepackage{appendix}

\begin{document}

\title{(Thermo-)dynamics of the spin-boson model in the weak coupling regime: Application as a quantum battery}

\author{Mahima Yadav}
\email{yadav.15@iitj.ac.in}
\author{Devvrat Tiwari}
\email{devvrat.1@iitj.ac.in}
\author{Subhashish Banerjee}
\email{subhashish@iitj.ac.in}
\affiliation{Indian Institute of Technology Jodhpur-342030, India\textsuperscript{}}
\date{\today}

\begin{abstract}
    We investigate the spin-boson model's dynamical and thermodynamic features in the weak coupling regime using the weak coupling spin-boson (WCSB) and phase covariant (PC) master equations. Both unital (pure dephasing) and non-unital (dissipative) quantum channels are considered. On the dynamical side, we explore key quantum features including non-Markovianity, quantum speed limit, quantum coherence, and the system’s steady-state behavior. Notably, the measures of non-Markovianity exhibit different behavior under WCSB and PC dynamics. From the quantum thermodynamic perspective, we conceptualize the spin-boson system as a quantum battery and analyze its performance through metrics such as energy, ergotropy, anti-ergotropy, and battery capacity. We further examine the roles of pure dephasing and dissipative processes in shaping the battery's performance. Our findings demonstrate the spin-boson model’s versatility as a platform for efficient energy storage and transfer in quantum thermodynamic devices.
\end{abstract}

\maketitle

\section{Introduction}\label{intro}
A quantum system inevitably interacts with its ambient environment. These interactions significantly impact its evolution. A framework for studying them is provided by the theory of open quantum systems ~\cite{Weiss2011, Breuer_book, SB_book, Louisell1973, CALDEIRA1983, GRABERT1988, Gardiner2004}. Recent years have seen significant progress in this subject and have been utilized to study a variety of problems~\cite{SGAD_channel, Tanimura_2020, Hughes_2009, Nazir_2014, SB_2003, Huelga_2013, Hanggi_2020, Tiwari_chaos, tiwari2024_circuit, Bose_2024}. The evolution of an open quantum system can be examined using the Gorini-Kossakowski-Sudarshan-Lindblad (GKSL) master equation~\cite{GKLSpaper, Lindblad1976}, which makes use of Born-Markov and rotating wave approximations, modeling a Markovian dynamics. However, with advancements in theory and technology, rapid inroads have been made in the challenging domain of non-Markovian dynamics ~\cite{ vega_alonso, Hall_andersson_2014, Rivas_2014, CHRUSCINSKI20221, Utagi2020, Zaccone_2018, kading2025}. Numerous applications of non-Markovian dynamics have been explored, with significant research devoted to understanding and utilizing these phenomena~\cite{Pradeep_Kumar_NM, RHP2010, Shrikant_2018, Breuer2009, Thapliyal2017, Kumar2018, Rivas_2011, HeatEngineSB, Zaccone_2024, Javid_2019}.

A very prominent model in open quantum systems is the paradigmatic spin-boson model, describing an interaction between a two-level system and a bosonic bath~\cite{Garg_1985, Weiss2011, spin_boson_model_leggett, Larson, spin_boson_model_leggett, CALDEIRA1983, Breuer_book, SB_book}. In realistic scenarios, environmental coupling induces dissipation and decoherence, necessitating an open-system framework. The two-level system, central to quantum mechanics, models binary quantum states across diverse platforms, from spin-$1/2$ particles to photon polarization. This versatile model underpins the study of decoherence, energy transfer, and non-Markovian behavior in quantum systems~\cite{Weiss2011}. Recently, various techniques have been applied to the spin-boson model~\cite{Tanimura_2020, spin_boson_tech_7, spin_boson_tech_1, spin_boson_tech_2, spin_boson_tech_3, spin_boson_tech_4, spin_boson_tech_5, spin_boson_tech_6}. Thus, for example, in the weak coupling regime, an analytical form of a master equation describing the evolution of the spin-boson model using the time-convolutionless (TCL) technique has been obtained~\cite{Goan2011-ke, Goan_2010, Haase2018, TCL_method, Breuer2001}. 

The spin-boson model, in the weak coupling regime, leads to a weak coupling spin boson master equation (WCSB), and further, upon making the secular approximation, it leads to an interesting class of dynamics, {\it viz.} the phase covariant (PC) dynamics~\cite{Haase2018}. A PC map describes the physical processes involving absorption, emission, and pure dephasing and has been studied in a number of works~\cite{PC_master_3, Baruah2023, Filippov2020, Teittinen2018, PC_master_1, Lankinen_2016, PC_master_2}. Further, based on the angle between the coupling operator in the $xz-$plane and the $x-$axis, the spin-boson model's dynamical map can be used as a platform to study the impact of both (non-)unital channels on the dynamics of a quantum system~\cite{Haase2018}. 

The dynamical behavior of the spin-boson model captures several fundamental aspects of open quantum systems. A central feature is quantum coherence, which is inevitably degraded over time due to environmental coupling, yet its persistence is crucial for quantum information processing and coherent control~\cite{l1_norm, Coherence_Streltsov, Bhattacharya2018, Chitamber_coherence, Naikoo2019, Dixit2019, Landi_coherence}. Another interesting dynamical quantity is the quantum speed limit (QSL), which sets the minimal time required for the system to evolve between two distinguishable states~\cite{Mandelstam1991, MARGOLUS1998188, Mandelstam_book, Anandan_aharanov, Giovannetti_2003}. The QSL provides insights into the interplay between system-bath interactions and the speed of quantum evolution, often governed by both coherence and dissipation~\cite{Deffner_2017, PhysRevLett.111.010402, QSL_in_OQS, Aggarwal_2022, Mohan_2022, Paulson_2022,qsl_central_spin, qsl_wigner_2025, bouri2024, QSL_brachistochrone}. Over long timescales, the system typically relaxes to a steady state, which may or may not correspond to its thermal equilibrium state depending on the strength and nature of the coupling~\cite{Hanggi_2020, hmf_anders_review}. In the weak coupling limit, the steady state often aligns with the thermal state of the system~\cite{pathania2024}. These dynamical features offer a rich framework to study fundamental quantum processes.

Building on these dynamical foundations, the field of quantum thermodynamics seeks to extend the principles of classical thermodynamics into the quantum regime~\cite{Gemmer2009, Binder_book, deffner_book, kosloff_2022, Alicki_1979}. This framework studies how heat, work, and energy are defined and exchanged in a quantum system, especially when they interact with a bath~\cite{tiwari2024, Landi_2021_entropy, Landi_2022_thermal_current}. A particularly compelling application within this domain is the concept of a quantum battery~\cite{Alicki_battery, Binder_2015, Carrega_2020, andolina_oqs_battery, Riccardo_2024, Shastri2025}. It is a finite-sized quantum system designed to store and deliver energy to a quantum system~\cite{quantum_battery_review, Fischer_battery, tiwari_impact1, Tiwari_impact2, Cavaliere2025, GRAZI2025}. The quantum battery has been realized using various platforms, including extended Dicke quantum battery~\cite{extended_dicke_battery}, self-discharge mitigated quantum battery~\cite{self_discharge_mitigated_QB}, resonator-qutrit quantum battery~\cite{resonator-qutrit_QB}, and Rosen-Zener quantum battery~\cite{rosen-Zener_QB}.
The maximum amount of work that can be extracted from a quantum battery is bounded by the ergotropy~\cite{Allahverdyan_2004}. It was shown in~\cite{Francica_CoherentErgotropy} that the ergotropy can be divided into incoherent and coherent parts. A recent figure of merit, $viz.$, battery capacity together with anti-ergotropy was developed in~\cite{battery_capacity}. The battery capacity sheds light on the storage of work in a quantum battery, and the anti-ergotropy provides a bound on the charging capacity of the quantum battery. 

In this work, the WCSB and PC master equations are used to investigate the dynamical and thermodynamic behavior of the spin-boson model. The spin-boson model has rich dynamic features, and based on the different couplings, the dynamics can be pure dephasing or dissipative. Here, we shall investigate the memory effect, speed of evolution, coherence dynamics, and the steady state behavior of the spin-boson model. These explorations will unravel a deeper insight into the various regimes of the spin-boson model. The spin-boson model is widely applicable in many physical realizations of quantum systems, making it pertinent for studying it in the context of a quantum battery. To this end, we shall envisage the spin (qubit) system of the spin-boson model as a quantum battery and the bosonic bath as a charger or dissipator. Our objective is to find out how the spin-boson model performs as a quantum battery in dissipative or pure dephasing regimes. The maximum amount of work that can be extracted is characterized by the ergotropy of the system, along with its coherent and incoherent parts. An important aspect of the spin-boson quantum battery realization presented here is how the type of interaction (dissipative or pure dephasing) impacts the charging capacity and work storage capacity of the battery, which are characterized by the anti-ergotropy and battery capacity, respectively. Thus, apart from the interesting dynamical features of the spin-boson model, we also aim to deliver a comprehensive analysis of the spin-boson quantum battery using quantifiers like energy, power, ergotropy, anti-ergotropy, and battery capacity.

The plan of the paper is as follows. In Sec.~\ref{spin-boson-model}, we introduce the spin-boson model and the corresponding WCSB and PC master equations. The dynamical properties of the model, including non-Markovianity, quantum speed limit, coherence, and steady state, are studied in Sec.~\ref{non_Markovian_behavior}. Section~\ref{sec_quantum_thermo} discusses the quantum thermodynamic performance of the system, considered as a quantum battery. The conclusions are presented in Sec.~\ref{conclusions}.

\section{The spin-boson model}\label{spin-boson-model}
The Hamiltonian of the spin-boson model~\cite{spin_boson_model_leggett} is given by
\begin{align}
    H &= H_S + H_B + V, 
    \label{total_hamiltonian}
\end{align}
where $H_S$, $H_B$, and $V$ are the system, bath, and the interaction Hamiltonian, respectively, which are given by (for $\hbar = 1$)
\begin{align}
    H_S &= \frac{\omega_0}{2}\sigma_z, \nonumber \\
    H_B &= \sum_n \omega_n b_n^\dagger b_n, ~~\text{and}\nonumber\\
    V &= \left(\cos(\theta)\frac{\sigma_x}{2} + \sin(\theta)\frac{\sigma_z}{2}\right)\sum_n(g_n b_n + g_{n}^{*}b_n^\dagger).
\end{align}
Here, $\omega_0$ is the transition frequency of the two-level atomic system, $b_n~(b_n^\dagger)$ are the bosonic annihilation (creation) operators corresponding to the bath mode $n$ with frequency $\omega_n$, $g_n$ is the strength of interaction between the system and the bath. The angle between the $x-$axis and the direction of the coupling operator in the $xz$-plane is given by the parameter $\theta$, such that the angle $\theta = \frac{\pi}{2}$ is for pure dephasing interaction~\cite{Breuer_book, SB_book} and $\theta = 0$ is for pure transversal (or perpendicular) interaction~\cite{Breuer_2012}. The above Hamiltonian is equal to the standard spin-boson model Hamiltonian~\cite{Weiss2011} modulo a unitary transformation~\cite{Haase2018}.
The dynamics of the spin system considering the bath to be initially in thermal state $\rho_B(0) = \exp\left(-\beta H_B\right)/{\rm Tr}\left[\exp\left(-\beta H\right)\right]$, ($\beta = \left(k_B T\right)^{-1}$ is the inverse temperature) was obtained in~\cite{Haase2018} using the time-convolutionless (TCL) master equation~\cite{Shibata1977, Chaturvedi1979, Breuer2001} up to the second order considering the bath to be weakly coupled to the system. The corresponding weak-coupling spin-boson (WCSB) master equation was shown to be
\begin{align}
    \frac{d\rho_S(t)}{dt} &= \mathcal{L}\left[\rho_S(t)\right] \nonumber \\
    &= -i\left[\tilde H_S(t), \rho_S(t)\right] \nonumber \\
    & + \sum_{j, k = \pm, z} \gamma_{kj} (t)\left(\sigma_k\rho_S(t)\sigma_j^\dagger - \frac{1}{2}\left\{\sigma_j^\dagger\sigma_k, \rho_S(t)\right\}\right),
    \label{gen_spin-boson_master_eq}
\end{align}
where 
\begin{align}
    \gamma_{zz}(t) &= \frac{\sin^2(\theta)}{2}\Re \left\{\xi(0, t)\right\},  \nonumber \\
    \gamma_{++}(t) &= \frac{\cos^2(\theta)}{2}\Re \left\{\xi(-\omega_0, t)\right\},  \nonumber \\
    \gamma_{--}(t) &= \frac{\cos^2(\theta)}{2}\Re \left\{\xi(\omega_0, t)\right\},  \nonumber \\
    \gamma_{+-}(t) &= \gamma_{-+}^{*}(t) = \frac{\cos^2(\theta)}{4} \left[ \xi(-\omega_0, t) + \xi^{*}(\omega_0, t) \right],  \nonumber \\
    \gamma_{z+}(t) &= \gamma_{+z}^{*}(t) = \frac{\sin{\theta} \cos{\theta}}{4}\left[\xi(0, t) + \xi^{*}(-\omega_0, t)\right],  \nonumber \\
    \gamma_{z-}(t) &= \gamma_{-z}^{*}(t) = \frac{\sin{\theta} \cos{\theta}}{4}\left[\xi(0, t) + \xi^{*}(\omega_0, t)\right].
    \label{init_rates}
\end{align}
$\Re\left\{z\right\}$ and $\Im\left\{z\right\}$ denotes the real and imaginary parts of $z$, respectively.  
The factor $\xi(\zeta, t) = \int_0^t ds ~e^{i\zeta s} C(s)$, where 
\begin{align}
    C(t) = \int_0^\infty d\omega~ J(\omega) \left[\coth\left(\frac{\omega}{2T}\right)\cos(\omega t) - i \sin (\omega t)\right],
    \label{C_S}
\end{align}
is the bath correlation function and $J(\omega) = \sum_n g_n^2\delta(\omega - \omega_n)$ is the spectral density of the bath. Further, the term $\tilde H_S(t)$ in Eq.~\eqref{gen_spin-boson_master_eq} can be written as
\begin{align}
    \tilde H_S(t) = H_S + H^{LS}(t),
\end{align}
where the time-dependent factor $H^{LS}(t)$ is the Hamiltonian contribution emerging because of the system-bath interaction, and whose matrix form is 
\begin{equation}
    H^{LS}(t) = \begin{pmatrix}
        H^{LS}_{00} & H^{LS}_{01}\\ \\
        H^{LS*}_{01} & H^{LS}_{11}
    \end{pmatrix},
\end{equation}
with $H^{LS}_{00} = \frac{\cos^2(\theta)}{4}\Im\left\{\xi(\omega_0, t)\right\}$, $H^{LS}_{11} = \frac{\cos^2(\theta)}{4}\Im\left\{\xi(-\omega_0, t)\right\}$, and $H^{LS}_{01} = -i\frac{\cos \theta \sin \theta}{4} 
\left[ \Re \{\xi(0, t)\} - \frac{1}{2} \left\{ \xi^*(-\omega_0, t) + \xi(\omega_0, t) \right\} \right]$. 
It is equivalent to the master equation derived in~\cite{Goan_2010, Goan2011-ke}, see Appendix B.

Throughout this paper, we consider the Ohmic bath spectral density with the Lorentz-Drude cutoff function given by
\begin{align}
    J(\omega) = \frac{2  m  \gamma}{\pi}  \omega \frac{\Omega^2}{\Omega^2 + \omega^2},
    \label{drude-Lorentz_spec_dens}
\end{align}
where $\Omega$ is the high-frequency cutoff, $m$ is the effective mass of bath's oscillators and $\gamma$ is the damping rate related to the system-bath coupling. The above form of the spectral density can be substituted in Eq.~\eqref{C_S} to obtain (for $\hbar = k_B = 1$)
\begin{align}
    \Re\{C(t)\} = 2m\gamma\Omega^2T\left[\frac{e^{-\Omega t}}{\Omega} + 2\sum_{n = 1}^\infty \frac{\Omega e^{-\Omega t} - \nu_n e^{-\nu_n t}}{\Omega^2 - \nu_n^2}\right], 
    \label{re_Ct}
\end{align}
where $\nu_n = 2\pi n T$ are the well-known Matsubara frequencies. Further, the imaginary part of $C(t)$ is 
\begin{align}
    \Im\{C(t)\} = -m\gamma\Omega^2e^{-\Omega t}.
    \label{im_Ct}
\end{align}
Now, the rates $\gamma_{kj}$ in Eq.~\eqref{init_rates} and $H^{LS}(t)$ contain the factors $\xi(0, t)$ and $\xi(\pm \omega_0, t)$, which can be analytically calculated for the spectral density, Eq.~\eqref{drude-Lorentz_spec_dens}. The forms of these functions are given in Appendix~\ref{appendix_A}. Note that the Markov approximation~\cite{Breuer_book} has not been applied in the derivation of the above master equation. The weak coupling approximation made here implies the neglect of the higher-order terms in the TCL expansion. The second-order TCL approximation is valid in the weak coupling regime of the spin-boson model. The validity of the second-order TCL approximation, particularly for the Ohmic bath spectral density with the Lorentz-Drude cutoff, depends on the appropriate limits of the damping factor $\gamma$ related to system-bath coupling, temperature $T$, and the system frequency $\omega_0$, see~\cite{Breuer2001}. For example, for low temperatures, the approximation is good if $\gamma$ is very small compared to the system frequency $\omega_0$. As we increase the temperature, the factor $\gamma$ can be slightly increased, though it must be small compared to $\omega_0$. This is a limitation of the TCL approximation, as it is not valid in the strong coupling regime. The advantage of using the TCL approximation is that it is applicable to a wide class of physical systems. It doesn't depend on the form of spectral density, the type of interaction, or the initial state of the combined system. Further, the TCL approximation provides a way to go beyond the Markovian approximation, which is shown in the subsequent section. As the coupling gets weaker, particularly for $\gamma\ll\omega_0$, the evolution tends to be Markovian, see Fig.~\ref{fig_RHPMeasure_wcsb} below.

Further, the pure dephasing as well as dissipative regimes of the master equation, Eq.~\eqref{gen_spin-boson_master_eq}, can be arrived at for certain values of the angle $\theta$. Specifically, if we put $\theta = \pi/2$, all the terms except the one containing $\gamma_{zz}$ vanish depicting the pure dephasing evolution, and if $\theta = 0$, only the terms containing $\gamma_{++}, \gamma_{+-}, \gamma_{--}$, and $\gamma_{-+}$ remain, denoting a dissipative evolution. 

Further simplification of the above master equation, Eq.~\eqref{gen_spin-boson_master_eq}, can be made by applying the secular approximation~\cite{Breuer_book, Breuer2001, Maniscalco2004, Fleming2010}, based on the system's evolution timescale $\tau_0$ and the relaxation time $\tau_R$ due to the interaction with the bath. Under the assumption $\tau_0\ll\tau_R$, the highly oscillatory terms can be averaged out to be zero, and the master equation of the system reduces to~\cite{Haase2018}
\begin{align}
    \frac{d\rho_S(t)}{dt} &= -i\left[H_S + \left(\frac{H_{11}^{LS}(t)}{2}\right)\sigma_z,  
      \rho_S(t)\right] \nonumber \\
    &+  \gamma_{++}(t) \left(\sigma_+\rho_S(t)\sigma_- - \frac{1}{2}\left\{\sigma_-\sigma_+, \rho_S(t)\right\}\right) \nonumber \\
    & + \gamma_{--}(t) \left(\sigma_-\rho_S(t)\sigma_+ - \frac{1}{2}\left\{\sigma_+\sigma_-, \rho_S(t)\right\}\right) \nonumber \\
    & + \gamma_{zz}(t) \left[\sigma_z\rho_S(t)\sigma_z -  \rho_S(t) \right],
    \label{PC_dynamics}
\end{align}
where the forms of the coefficients $\gamma_{++}, \gamma_{--}$, and $\gamma_{zz}$ are given in Eq.~\eqref{init_rates}. This master equation corresponds to the most general form of a phase covariant (PC) master equation~\cite{Teittinen2018, Baruah2023, Filippov2020, Haase2018}. 

The system's dynamics, in this case, can be explicitly calculated. Consider the initial state of the system is given by
\begin{align}
    \rho_S(0) = \frac{1}{2}\begin{pmatrix}
        1 + r_z(0) & r_x(0) - i r_y(0) \\
        r_x(0) + i r_y(0) & 1 - r_z(0) 
    \end{pmatrix},
\end{align}
where $r_k(0) = {\rm Tr} \left[\sigma_k \rho_S(0)\right]$ (for $k = x, y, z$), where $\sigma_{k}$'s are the Pauli matrices, which define the components of the Bloch vector. The solution of the phase covariant master equation, Eq.~\eqref{PC_dynamics}, is given by~\cite{Teittinen2018, Lankinen_2016} 
\begin{align}
\rho_S(t) = 
\frac{1}{2}\begin{pmatrix}
        1 + r_z(t) & r_x(t) - i r_y(t) \\
        r_x(t) + i r_y(t) & 1 - r_z(t) 
    \end{pmatrix},
\end{align}
where
\begin{align}
    r_x(t) &= r_x(0) \exp\left[i \chi(t) - \frac{\Gamma (t)}{2} - \widetilde \Gamma(t)\right],  \nonumber \\
    r_y(t) &= r_y(0) \exp\left[i \chi(t) - \frac{\Gamma (t)}{2} - \widetilde \Gamma(t)\right],
\end{align}
and 
\begin{align}
    r_z(t) &= 1 - \exp[-\Gamma(t)]\left[2G(t) + 1 - r_z(0)\right].
\end{align}
The factors $G(t), \Gamma(t), \widetilde \Gamma(t),$ and $\chi(t)$ in the expressions of $x(t), y(t)$, and $z(t)$ are given by 
\begin{align}
    G(t) &= \int_0^t e^{\Gamma(s)}\gamma_{--}(s)~ds, \nonumber \\
    \Gamma(t) &= \int_0^t \left[\gamma_{++}(s) + \gamma_{--}(s)\right]~ds, \nonumber \\
    \widetilde \Gamma(t) &= \int_0^t 2\gamma_{zz}(s)~ds,~~\text{and} \nonumber \\
    \chi(t) &= \int_0^t \left[2\omega_0 + \frac{\cos^2(\theta)}{2} \Im \left\{\xi \left(-\omega_0, s\right)\right\}\right]~ds.
\end{align}

In the subsequent sections, we study the dynamical and thermodynamical properties of the system for both unital (pure dephasing evolution) and non-unital (dissipative evolution) cases, taking into account the WCSB [Eq.~\eqref{gen_spin-boson_master_eq}] and PC [Eq.~\eqref{PC_dynamics}] master equations, governing the evolution of the system.

\section{Non-Markovianity, speed limit, quantum coherence, and steady state of the system}\label{non_Markovian_behavior}
Here, we study the dynamical aspects of the spin-boson model, governed by the WCSB and PC master equations, Eqs.~\eqref{gen_spin-boson_master_eq} and~\eqref{PC_dynamics}, respectively. The non-Markovianity of the system is investigated. Further, we study the quantum speed limits of the spin-boson model, which is an important facet of a quantum system's dynamics. Also, we find out the changes in the coherence of the spin system during its evolution. The steady state of the system is studied to shed light on its thermal equilibrium properties. 

\subsection{Non-Markovianity}
To investigate the non-Markovian behavior of the system, we use two measures of non-Markovianity, which are commonly known as the trace distance measure or the Breuer-Laine-Piilo (BLP) measure~\cite{Breuer2009}, and the Rivas, Huelga, and Plenio (RHP) measure~\cite{RHP2010, Rivas_2014}. 
\subsubsection{Trace distance measure}
Consider two quantum states $\rho_1$ and $\rho_2$, the trace distance, a measure of distinguishability between the quantum states, is given by
\begin{equation}
    \mathcal{D}(\rho_{1}, \rho_{2}) = \frac{1}{2} {\rm Tr}\left| \rho_{1} - \rho_{2} \right|,
\end{equation}
where $ |A| = \sqrt{A^\dagger A}$. In~\cite{Breuer2009}, it was argued that a dynamical decrease of $\mathcal{D}(\rho_{1}, \rho_{2})$ can be attributed to a loss of information from the open
system into the environment. This is a characteristic of Markovian dynamics. However, a revival in the variation of trace distance indicates a flow of information from the environment back to the system, indicating non-Markovian behavior.
The two initial states of the reduced system we use in this analysis are given by
\begin{align}
    \rho_{1}(0) = \begin{pmatrix}
        1/2  & 1/2 \\
        1/2 & 1/2   
    \end{pmatrix}, && \text{and} &&
    \rho_{2}(0) = \begin{pmatrix}
        1 & 0 \\
        0 & 0   
    \end{pmatrix}.
\end{align}

\begin{figure}
    \centering
    \includegraphics[width=1\linewidth]{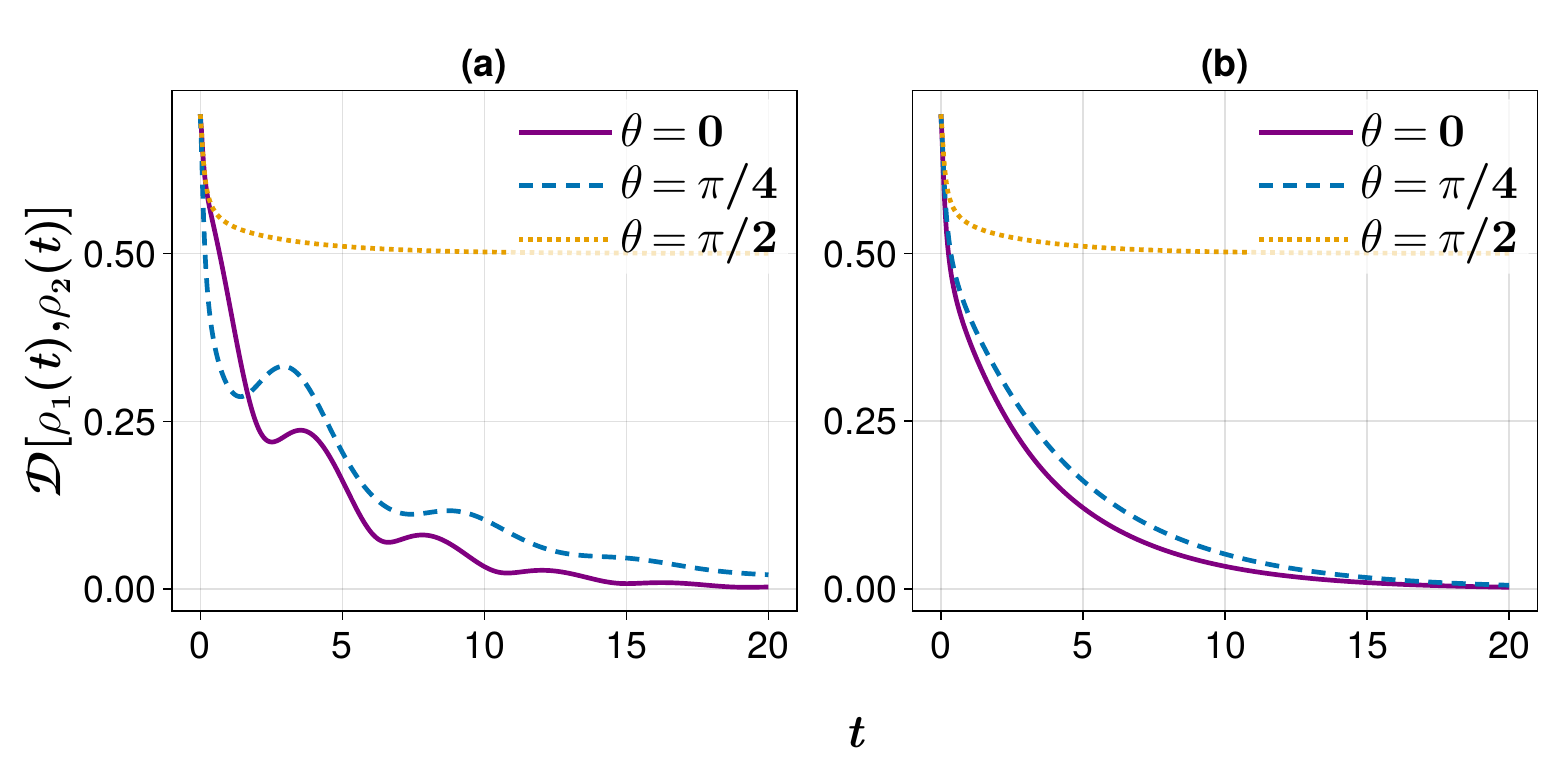}
    \caption{Variation of trace distance $\mathcal{D}\left[\rho_{1}(t), \rho_{2}(t)\right]$ with time for different coupling angles $\theta$. The evolution of the system is governed by the WCSB master equation~\eqref{gen_spin-boson_master_eq} in (a), and the PC master equation~\eqref{PC_dynamics} in (b). The parameters are taken to be $\omega_0 = 1.25$, $T= 0.2$, $\Omega = 15$, $m\gamma = 0.4$.}
    \label{BLPMeasure}
\end{figure}

In Fig.~\ref{BLPMeasure}(a), we plot the variation of trace distance with time for the system's evolution governed by Eq.~\eqref{gen_spin-boson_master_eq}. The behavior of the trace distance provides insights into the system's (non-)Markovian evolution. For \(\theta = \pi/2\), the trace distance decreases monotonically over time, indicating the absence of information backflow associated with pure dephasing (unital evolution). This indicates that the evolution of the system, for the $\theta = \pi/2$ case, is BLP Markovian. In contrast, for \(\theta = 0\), the trace distance exhibits oscillatory behavior due to dissipative interactions, which are inherently non-unital. As the evolution progresses, these interactions cause the trace distance to vanish. For \(\theta = \pi/4\), where the system undergoes a mixed evolution, the trace distance decreases while displaying oscillations, signaling the presence of non-Markovian dynamics.
Furthermore, when the evolution is restricted to the phase covariant case, Eq.~\eqref{PC_dynamics}, the trace distance remains monotonic for all values of \(\theta\), showing non-oscillatory behavior, making this a $P$-divisible process, as can be observed from Fig.~\ref{BLPMeasure}(b). However, the evolution is still non-Markovian, as evidenced by the RHP measure, discussed below.

\subsubsection{RHP Measure}
The BLP measure quantifies non-Markovianity based on the behavior of the trace distance and effectively witnesses non-Markovian dynamics. However, it may fail to detect deviations from Markovianity in certain cases~\cite{Utagi2020}, particularly when non-Markovian effects manifest solely through coherence revivals or just through the non-unital part of dynamics, as it relies exclusively on the trace distance between quantum states. Therefore, to analyze the (non-)Markovian dynamics of a phase-covariant channel, which encompasses both unital and non-unital evolutions, we employ the RHP measure.

The RHP measure of non-Markovianity, introduced in~\cite{RHP2010, Rivas_2014}, is based on deviations from the complete positivity (CP) of the intermediate dynamics. The dynamical map $\mathcal{E}_{\left(t, t_0\right)} = \mathcal{E}_{\left(t, t_1\right)} \mathcal{E}_{\left(t_1,  t_0\right)}$ is CP-divisible if the intermediate map $\mathcal{E}_{\left(t, t_1\right)}$ is positive at all times $t_1$. Consequently, the evolution would be non-Markovian if the complete positivity condition is not satisfied for an intermediate map. 

For a map $\mathcal{E}_{\left(t, t_1\right)}$ to be completely positive, it must satisfy the positive semidefinite condition. Additionally, based on its trace-preserving property, a function \( g(t) \) is defined as 
\begin{equation}
    g(t) = \lim_{\epsilon \to 0^+} \frac{\| [ \mathcal{E}_{\left(t + \epsilon, t\right)} \otimes \mathbb{I})] \left( \vert\Phi\rangle \langle \Phi\vert \right)\|_{1} - 1}{\epsilon},
\end{equation}
where $\vert\Phi\rangle = \frac{1}{\sqrt{d}} \sum_{n=0}^{d-1}\vert n\rangle \vert n\rangle $  is maximally entangled state and $\| A\|_{1} = {\rm Tr}\sqrt{A^{\dagger}A}$ is the trace norm of matrix $A$. \( g(t) > 0 \) indicates non-Markovian evolution at time \( t \). 
For the generator $\mathcal{L}_{t}$ of the master equation, Eq.~\eqref{PC_dynamics}, the function $g(t)$ is defined as
\begin{equation}\label{eq_function_gt}
    g(t) = \lim_{\epsilon \to 0^+} \frac{\| [ \mathbb{I} + \epsilon(\mathcal{L}_{t} \otimes \mathbb{I})] \left( \vert\Phi\rangle \langle \Phi\vert \right)\|_{1} - 1}{\epsilon}.
\end{equation}  

\begin{figure}
    \centering
    \includegraphics[width=1\linewidth]{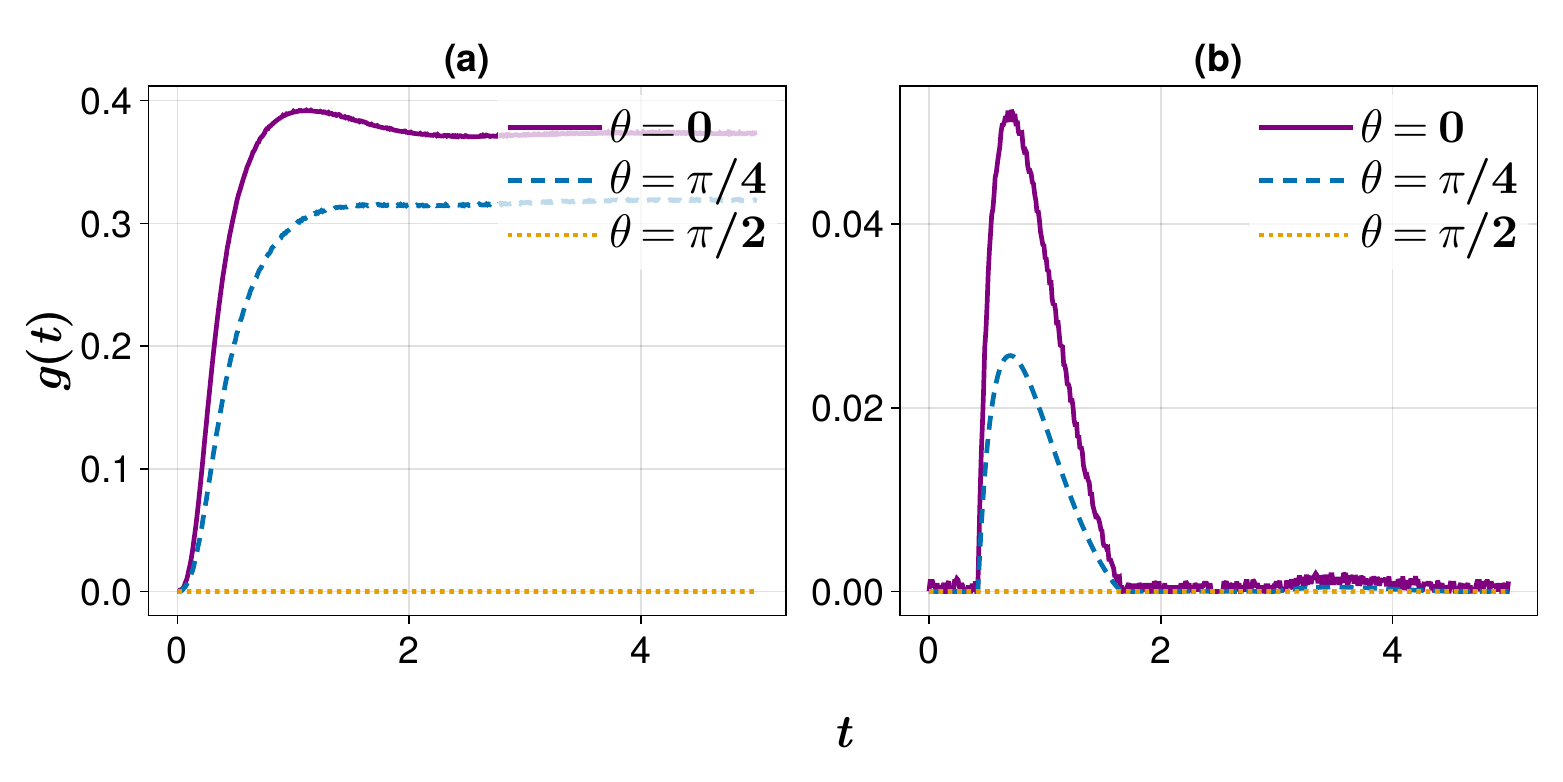}
    \caption{Variation of the function $g(t)$, Eq.~\eqref{eq_function_gt}, with time for different coupling angles $\theta$. The evolution of the system is governed by the WCSB master equation~\eqref{gen_spin-boson_master_eq} in (a), and the PC master equation~\eqref{PC_dynamics} in (b). Here, we have $\omega_0 = 2.25$, $T= 0.2$, $\Omega = 15$, $m\gamma = 0.4$.}
    \label{RHPMeasure}
\end{figure}

The function $g(t)$ is illustrated in Fig.~\ref{RHPMeasure}. Figure~\ref{RHPMeasure}(a) depicts the evolution of $g(t)$ for the WCSB master equation, Eq.~\eqref{gen_spin-boson_master_eq}. This figure confirms that the dissipative dynamics under the WCSB master equation exhibit non-Markovian behavior, while the evolution is RHP Markovian in the case of pure dephasing dynamics, as $g(t) = 0$ when $\theta = \pi/2$.
To investigate the non-Markovianity of the phase covariant (PC) dynamics, the function \( g(t) \), calculated using Eq.~\eqref{PC_dynamics} in Eq.~\eqref{eq_function_gt}, is plotted in Fig.~\ref{RHPMeasure}(b).
We observe that \( g(t) \) takes positive values, indicating a deviation from Markovian dynamics for angles \( \theta = 0 \) and \( \theta = \pi/4 \). For \( \theta = \pi/2 \), which corresponds to unital dynamics, \( g(t) \) remains zero at all times, signifying a completely positive (CP) divisible map. However, as the contribution of the non-unital component increases, the deviation from CP divisibility (a signature of non-Markovianity) becomes more pronounced. In particular, for purely non-unital dynamics (\( \theta = 0 \)), the system exhibits the strongest non-Markovian behavior.
\begin{figure}
    \centering
    \includegraphics[width=1\linewidth]{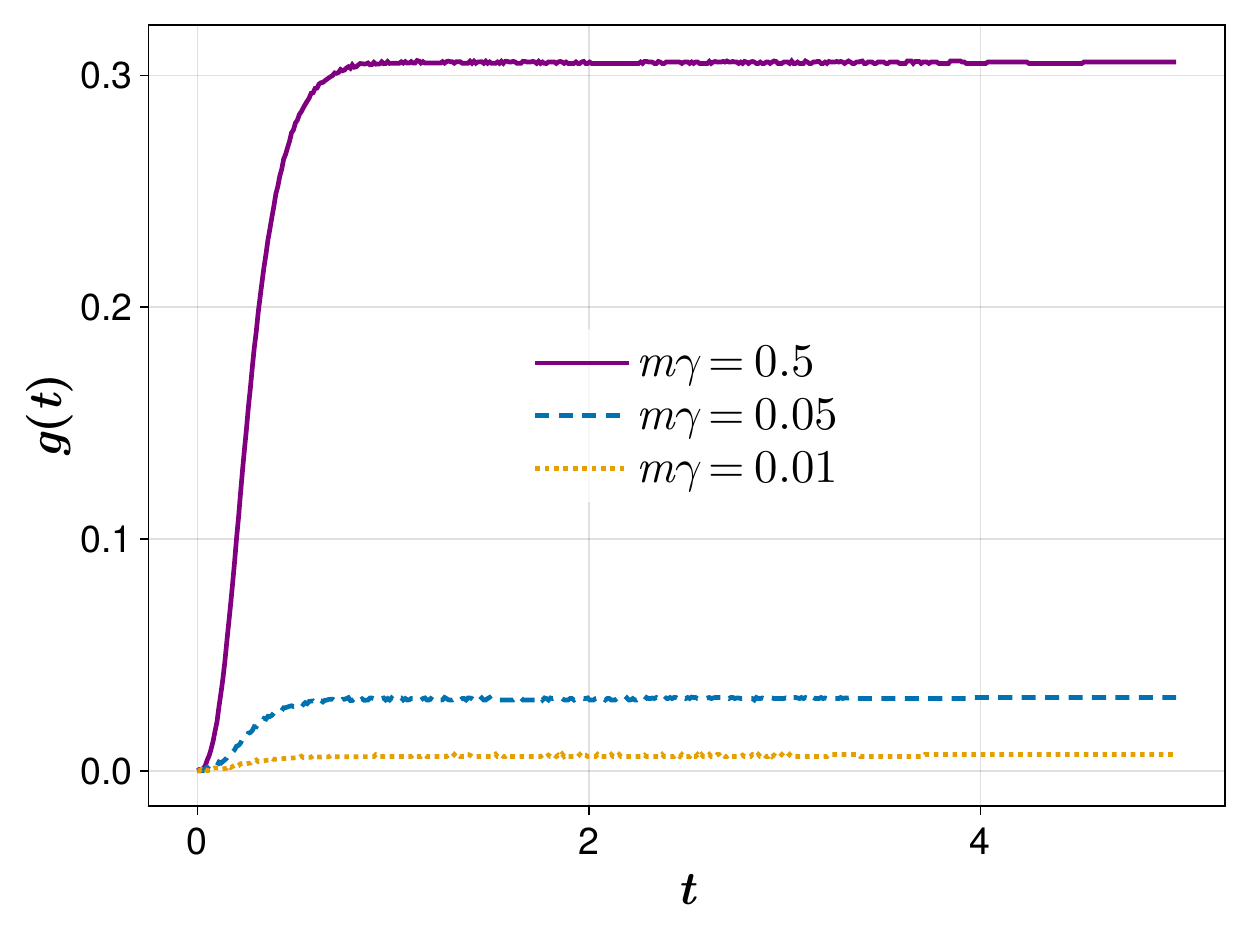}
    \caption{Variation of the function $g(t)$, Eq.~\eqref{eq_function_gt}, with time for different values of the coupling factor $\gamma$. The evolution of the system is governed by the WCSB master equation~\eqref{gen_spin-boson_master_eq}. Here, we take $\omega_0 = 2.25$, $T= 1.0$, $\Omega = 15$, and $\theta = 0$.}
    \label{fig_RHPMeasure_wcsb}
\end{figure}
Further, the function $g(t)$ is also plotted for different values of the system-bath coupling factor $\gamma$ in Fig.~\ref{fig_RHPMeasure_wcsb}. It can be observed that as we decrease the value of $\gamma$, the values of $g(t)$ diminish, depicting that as the coupling gets weaker, the system's evolution becomes less non-Markovian.

Thus, we can conclude that the evolution of the system using the PC master equation [Eq.~\eqref{PC_dynamics}] is BLP Markovian but RHP non-Markovian. In contrast, the dissipative WCSB evolution [Eq.~\eqref{gen_spin-boson_master_eq}] is both BLP as well as RHP non-Markovian.

\subsection{Quantum speed limit of the evolution}
One of the prominent dynamic properties of a quantum system is its speed of evolution, which is generally studied using the quantum speed limit. Here, we determine the speed of evolution of the spin-boson model using the geometric quantum speed limit (QSL) time employing the Wigner–Yanase (WY) information metric. Quantum speed limit time $\tau_{QSL}$ ~\cite{Mandelstam1991, MARGOLUS1998188} using Wigner-Yanase (WY) metric~\cite{PhysRevLett.111.010402, Pires_2016} is given by 
\begin{equation}
    \tau_{QSL} = \max \left\{ \frac{1}{\Lambda_{t}^{\rm op}}, \frac{1}{\Lambda_{t}^{\rm tr}}, \frac{1}{\Lambda_{t}^{\rm hs}} \right\}\sin^2{\mathcal{B}},
\end{equation}
where 
\begin{equation}
    \Lambda_{t}^{\rm op,tr,hs} = \frac{1}{t}\int_{0}^t ds || \mathcal{L}(\rho_s)||_{\rm op,tr,hs},
\end{equation}
where the $\|(.)\|_{\rm op, tr, hs}$ denote the operator, trace, and Hilbert-Schmidt norms, respectively. It is known that the operator norm provides a tighter bound~\cite{PhysRevLett.111.010402}. Further, $\mathcal{B}$ is given by
\begin{equation}
    \mathcal{B}\left[\rho(0), \rho(t)\right] = \arccos{\left({\rm Tr}\left[\sqrt{\rho(0)}\sqrt{\rho(t)}\right]\right)}, 
\end{equation}
where ${\rm Tr}\left[\sqrt{\rho(0)}\sqrt{\rho(t)}\right]$ is known as quantum affinity. It was shown in~\cite{Pires_2016, tiwari_impact1} that using the WY metric, one can obtain a tighter bound on the QSL time for the mixed quantum state evolution. 
\begin{figure}
    \centering
    \includegraphics[width=1\linewidth]{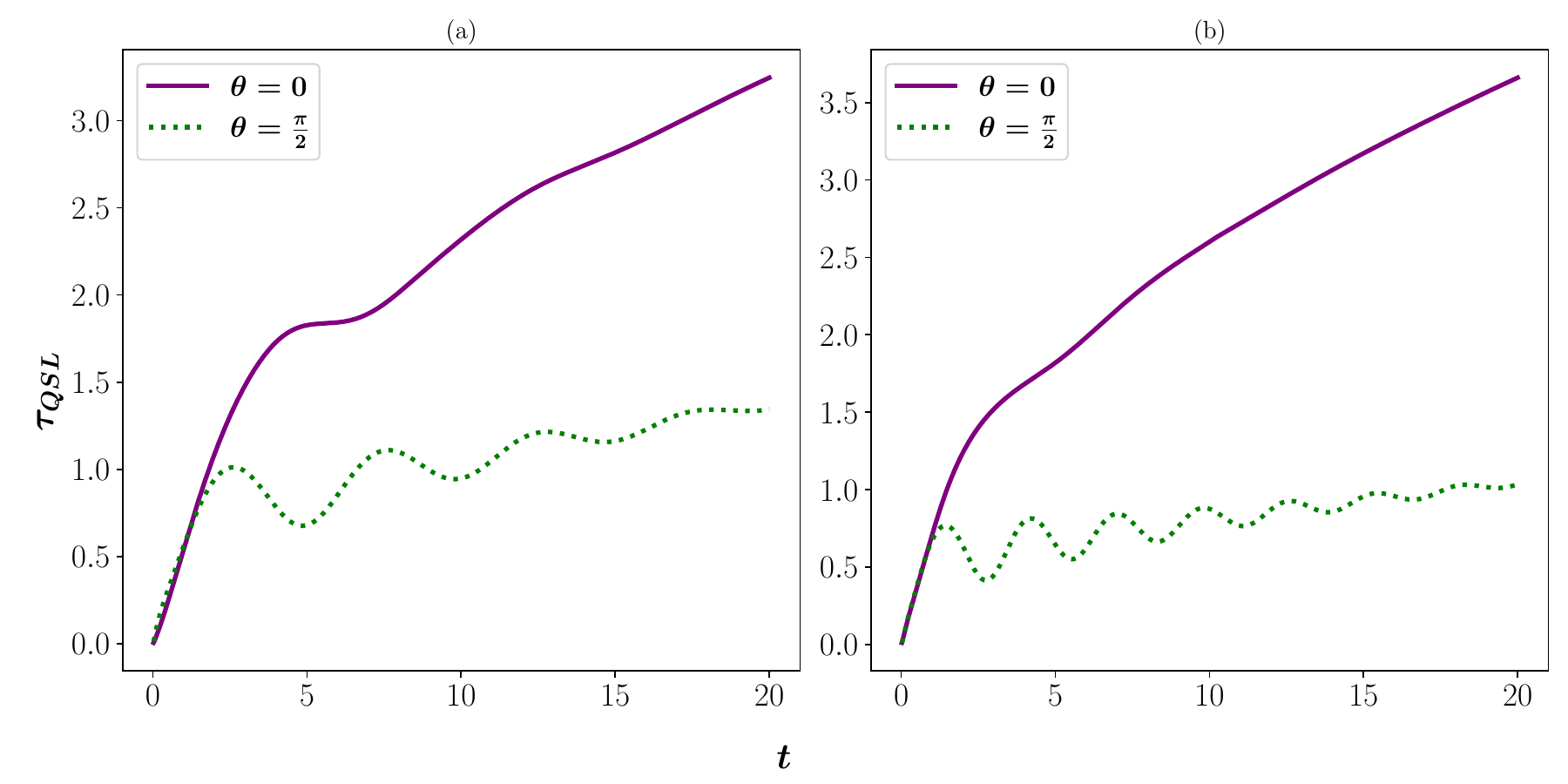}
    \caption{Variation of quantum speed limit time $\tau_{QSL}$ with time using WY Metric for different coupling angles $\theta$. In (a), the dynamics is governed by Eq.~\eqref{gen_spin-boson_master_eq}, and in (b), it is governed by the phase covariant master equation, Eq.~\eqref{PC_dynamics}. The parameters are $T = 0.2$, $\Omega = 15$, $m\gamma = 0.4$ and in (a) $\omega_0 = 1.25$ while in (b) $\omega_0 = 2.25$. The initial state of the system in both (a) and (b) is taken to be $\ket{\psi(0)}_S = \left(\frac{\sqrt{3}}{2}\ket{0} + \frac{1}{2}\ket{1}\right)$.}
    \label{fig_QSLv3}
\end{figure}
In Fig.~\ref{fig_QSLv3}, we plot the variation of the quantum speed limit time $\tau_{QSL}$ with the evolution time of the system. Further, Fig.~\ref{fig_QSLv3}(a) and (b) denote the evolution of the system via the general master equation, Eq.~\eqref{gen_spin-boson_master_eq}, and the phase covariant master equation, Eq.~\eqref{PC_dynamics}, respectively. It can be observed that the $\tau_{QSL}$ behaves in a similar manner for both cases. However, a stark difference is observed in the variation of $\tau_{QSL}$ for different values of the coupling angle $\theta$. 
Here, we observe that the speed of evolution is higher; that is, $\tau_{QSL}$ takes lower values when the evolution is purely dephasing, $\theta = \pi/2$, as compared to the dissipative case, $\theta = 0$. 

\subsection{Evolution of quantum coherence}
Quantum coherence is an important and unique feature of a quantum system. It is a key ingredient in present quantum technologies and provides a base for interesting quantum phenomena, such as entanglement. Recently, it has been studied as a physical resource~\cite{Coherence_Streltsov, Paulson_2022}. Here, we use the quantum relative entropy of coherence as a measure to study the evolution of coherence in the system. It is given by
\begin{align}
    C(\rho) = S(\rho^D) - S(\rho),
\end{align}
where \( \rho^D = \sum_i \langle i | \rho | i \rangle | i\rangle \langle i | \) is the dephased state obtained by removing the off-diagonal elements of \(\rho\). Here, \( S(\rho) = -\text{Tr}[\rho \log \rho] \) denotes the von Neumann entropy of \(\rho\).  
\begin{figure}
    \centering
    \includegraphics[width=0.95\linewidth]{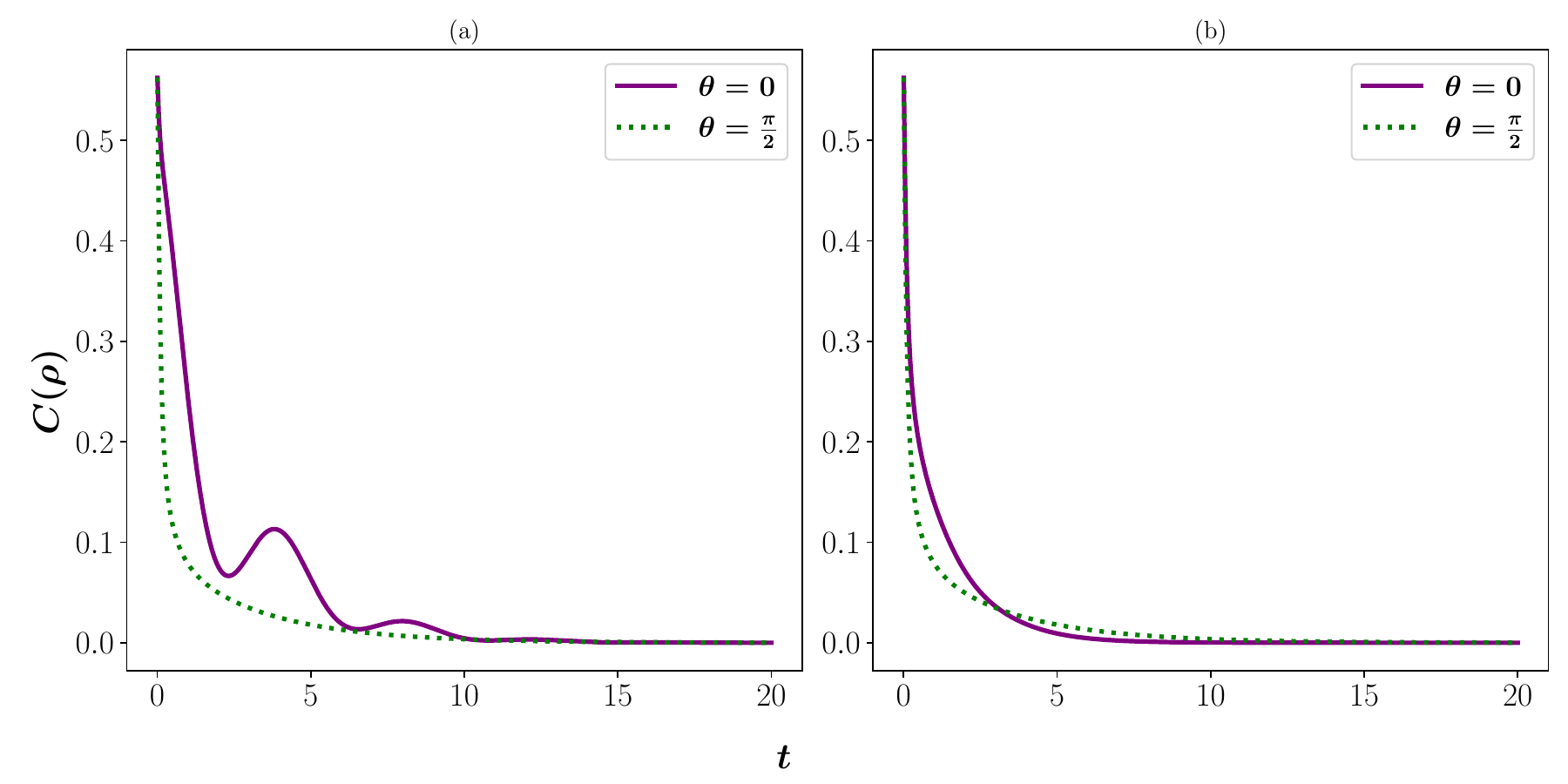}
    \caption{Variation of the quantum coherence with time for the evolution of the system using (a) the WCSB master equation, Eq.~\eqref{gen_spin-boson_master_eq}, and (b) the phase covariant master equation, Eq.~\eqref{PC_dynamics}. Parameters are temperature $ T = 0.2$, $\Omega = 15$, $m \gamma = 0.4$. In (a), $\omega_0 = 1.25$ and in (b), $\omega_0 = 2.25$.}
    \label{fig_CSLv1}
\end{figure}
We plot the variation of quantum coherence encompassing (non-)unital cases with the evolution governed by Eqs.~\eqref{gen_spin-boson_master_eq} (WCSB master equation) and~\eqref{PC_dynamics} (PC master equation) in Fig.~\ref{fig_CSLv1}. We observe that in both the WCSB and the PC dynamics, the behavior of the quantum coherence is the same in the case of pure dephasing evolution ($\theta = \pi/2$). Further, in the case of the PC dynamics, there is not a significant qualitative difference in the quantum coherence between the (non-)unital scenarios, as can be seen from Fig.~\ref{fig_CSLv1}(b). However, as can be observed from Fig.~\ref{fig_CSLv1}(a), when the dynamics is governed by the WCSB master equation, we observe that in the non-unital scenario ($\theta = 0$), quantum coherence revives multiple times after the initial decay. Further, in this case, quantum coherence takes higher values for $\theta = 0$ as compared to $\theta = \pi/2$.    

\subsection{Steady state of the system}
\begin{figure}
    \centering
    \includegraphics[width=1\linewidth]{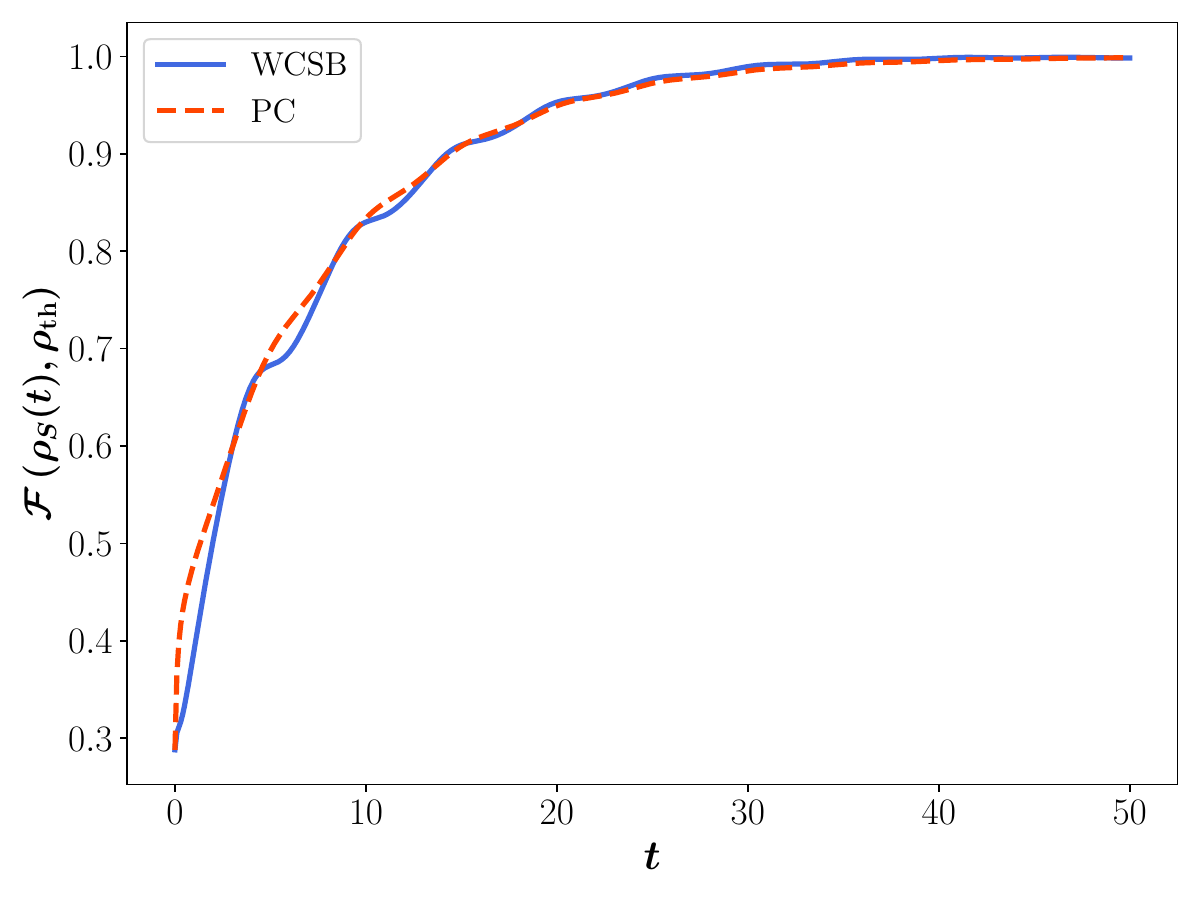}
    \caption{Variation of the fidelity $\mathcal{F}\left(\rho_S(t), \rho_{\rm th}\right)$ between the states evolved using WCSB [Eq.~\eqref{gen_spin-boson_master_eq}] and the PC [Eq.~\eqref{PC_dynamics}] master equations and the thermal state $\rho_{\rm th} = e^{-\beta H_S}/{\rm Tr}[e^{-\beta H_S}]$ corresponding to the system Hamiltonian $H_S$. The parameters are $ T = 0.5$, $\Omega = 15$, $m \gamma = 0.1, \theta = \pi/4$, and $\omega_0 = 1.25$.}
    \label{fig_SteadyStateTemp}
\end{figure}

A steady state of a quantum system provides important insights into the dynamics of the system. In the case of a single qubit system interacting with a bath, it is the system's state when the system has reached thermal equilibrium with the bath. The steady-state represents the long-term behavior of the system after transient effects have dissipated. The steady state can be obtained by setting the equation of motion $\frac{d\rho_S(t)}{dt}=0$. Another way to find out the system's steady state is to evolve the system to a long time limit; the corresponding state in this limit will be the system's steady state. We find the steady state of the system numerically when the system is evolved using the WCSB master equation, Eq.~\eqref{gen_spin-boson_master_eq}. In the case of the PC master equation, Eq.~\eqref{PC_dynamics}, derived from the spin-boson model, the form of the steady state can be analytically calculated, which is 
\begin{align}
    \rho_S^{SS} = \begin{pmatrix}
        \frac{\gamma_{++}}{\gamma_{--} + \gamma_{++}}&0\\
        0&\frac{\gamma_{--}}{\gamma_{--} + \gamma_{++}}
    \end{pmatrix},
\end{align}
for large $t$, where $\gamma_{--}(t)$ and $\gamma_{++}(t)$ become constant. To investigate the thermal equilibrium behavior of the system, we analyze the fidelity between the long-time evolved state $\rho_S(t)$, approaching the steady state, and the corresponding thermal equilibrium state $e^{-\beta H_S}/Z$, where $Z = {\rm Tr}\left[e^{-\beta H_S}\right]$, of the system. The corresponding fidelity, quantifying closeness between two quantum states, is given by
\begin{align}
    \mathcal{F}\left(\rho_S(t), \rho_{th}\right) = \left[{\rm Tr}\sqrt{\sqrt{\rho_{\rm th}}\rho_S(t)\sqrt{\rho_{\rm th}}}\right]^2.
\end{align}
In Fig.~\ref{fig_SteadyStateTemp}, we plot the variation of the fidelity $\mathcal{F}\left(\rho_S(t), \rho_{th}\right)$ with time. A comparison is made between the cases where the system's evolution is governed by the WCSB and the PC master equations. In the case of the system's evolution via the WCSB master equation, the fidelity has a slightly non-monotonic evolution, which smooths out in the PC dynamics. It can be observed that in the long-time limit, both cases lead the system's state to thermal equilibrium, indicated by fidelity attaining the value one. This brings out that the system reaches a thermal equilibrium state corresponding to its bare Hamiltonian in the long-time limit. This is consistent with the thermal equilibrium behavior of a quantum system in the weak coupling regime from the perspective of the Hamiltonian of mean force~\cite{pathania2024}.

\section{Quantum thermodynamics: Application as a quantum battery}\label{sec_quantum_thermo}
In thermodynamics, a fundamental question of interest is the exchange of energy between a system and its surroundings, as well as the efficiency with which the system can perform work. In quantum thermodynamics, phenomena such as entanglement, coherence, and decoherence introduce additional complexity. These quantum effects are particularly relevant when considering a quantum battery, a system designed to store and release energy in a controlled manner. Considering an open quantum system as a quantum battery makes it essential to study how system-bath interaction influences its performance in energy extraction and storage.
Here, we envisage the qubit system as a quantum battery interacting with the bosonic bath, which can act as a charger/dissipator. To characterize the performance of this quantum battery, we use quantifiers such as energy, ergotropy, and its (in-)coherent parts, instantaneous and charging powers, anti-ergotropy, and battery capacity. The initial state of the system is taken to be $\ket{\psi(0)}_S = \frac{\sqrt{3}}{2}\ket{0} + \frac{1}{2}\ket{1}$ throughout this section unless stated differently in the caption, where $\ket{0}(\ket{1})$ is excited (ground) state of the system. Further, we use the WCSB master equation in this section. The thermodynamic properties of the system when evolved using the PC master equation behave in a similar way, and are presented in Appendix C. 

\subsection{Energy and the instantaneous power}
The energy of the system at any time $t$ is given by $E(t) = {\rm Tr}\left[H_S\rho_S(t)\right]$. Using the Bloch vector representation of the system's state 
\begin{align}
    \rho_S(t) = \frac{1}{2} \left(\mathbb{I} + \sum_{k = x, y, z}k(t)\cdot\sigma_k\right),
    \label{eq_Bloch_vector_form}
\end{align}
where $k(t) = {\rm Tr}\left[\sigma_k\rho_S(t)\right]$ and $\mathbb{I}$ is the identity operator in system's subspace, 
we get the energy of the spin system to be
\begin{align}
    E(t) = \frac{\omega_0}{2}z(t).
\end{align}
The instantaneous power of the system is defined as the time derivative of the energy
\begin{align}
    P(t) = \frac{dE(t)}{dt} = \frac{\omega_0}{2}\dot z(t).
    \label{eq_inst_power}
\end{align}
\begin{figure}
    \centering
    \includegraphics[width=1\linewidth]{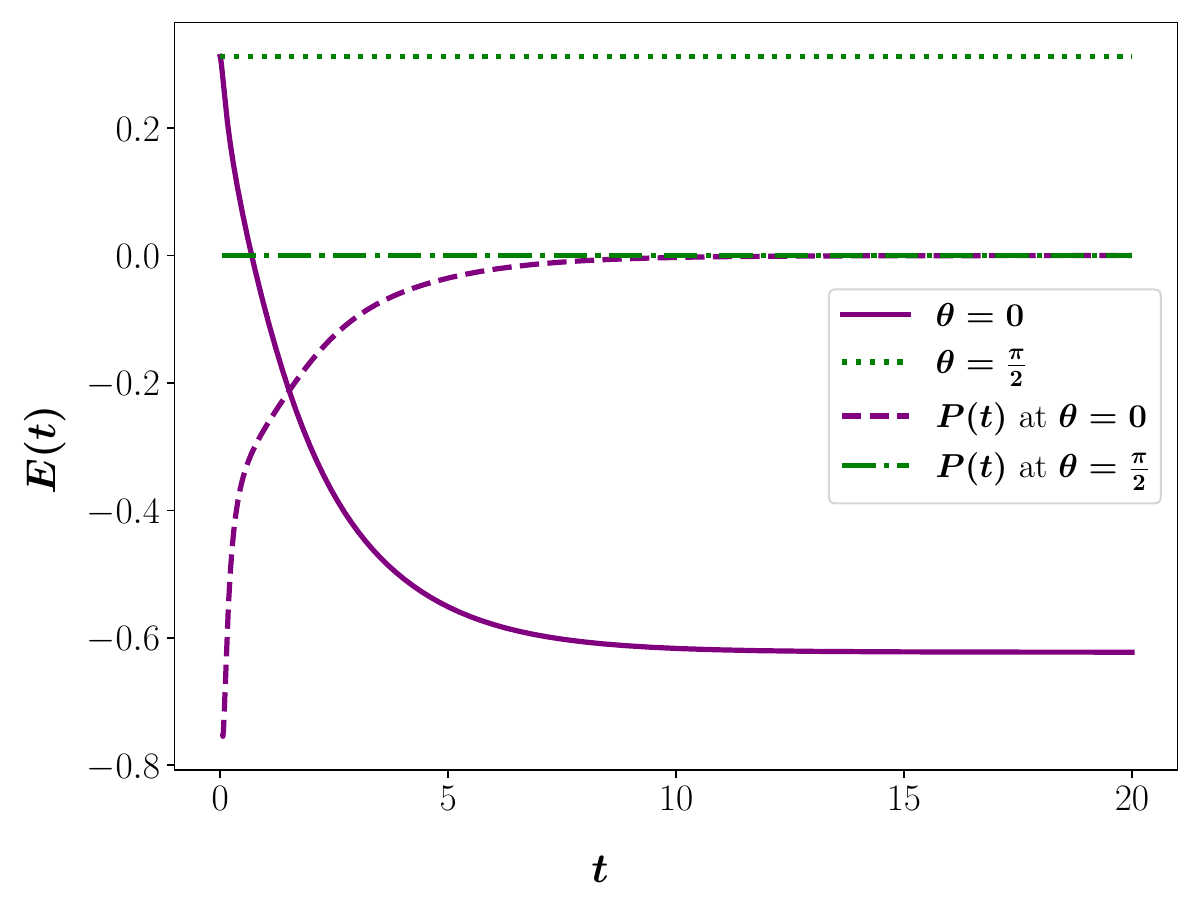}
    \caption{Variation of system's energy $E(t)$ and its instantaneous power $P(t)$ for (non-)unital evolution of the system governed by Eq.~\eqref{gen_spin-boson_master_eq}. The parameters are $\omega_0 = 1.25$, $T = 0.2$, $\Omega = 15$, and $m\gamma = 0.4$.}
    \label{Fig_EnergyPower1}
\end{figure}%
In Fig.~\ref{Fig_EnergyPower1}, we plot the variation of the system's energy and the instantaneous power at any time $t$ for both pure dephasing ($\theta = \pi/2$) and dissipative ($\theta = 0$) dynamics of the system governed by Eq.~\eqref{gen_spin-boson_master_eq}. For the pure dephasing case, we observe that the energy $E(t)$ remains constant because the population (diagonal) terms of the density matrix remain unchanged. As a result, the instantaneous power $P(t)$, which represents the rate of change of energy, is zero throughout the evolution. In contrast, for the dissipative case, energy dissipates and reaches a constant negative value. This renders the instantaneous power initially negative, and it becomes zero when the energy is constant. A negative instantaneous power, here, depicts the scenario where the system loses its internal energy to the environment.

\subsection{Ergotropy}\label{energy_ergotropy}
Ergotropy of a quantum system is a measure that provides an upper bound on the maximum extractable work from a quantum system via unitary transformations and also serves as an important quantifier for the charging and discharging behavior of a quantum battery~\cite{Allahverdyan_2004, Kamin_2020}. In the case of open quantum systems, ergotropy is calculated by considering the evolved state of the system as input to the ergotropy computation~\cite{Cakmak}. This allows us to determine the bound on work extraction via unitary transformations while accounting for environmental effects.
The ergotropy of a $d$-dimensional quantum system [governed by a Hamiltonian $H_S$, and whose state at any time $t$ is $\rho_S(t)$] is given by
\begin{align}
\mathcal{W}[\rho_S (t)] = \text{Tr} \left[\rho_S (t) H_S \right] - \text{Tr} \left[\rho_S^p H_S \right],
\end{align}
where \( \rho_S^p \) is the passive state corresponding to the input state \( \rho_S(t) \), which originally had the potential to perform work. After the system undergoes an optimal unitary evolution under a cyclic potential, it reaches this passive state, from which no further work can be extracted. Unlike macroscopic systems, where passive states often correspond to thermal Gibbs states, in quantum systems, the passive state is generally different.  
The spectral decomposition of the passive state is given by
\begin{align}
    \rho_S^p = \sum_{j=1}^{d} r_j |\epsilon_j\rangle \langle \epsilon_j|,
\end{align}
where the $\ket{\epsilon_j}$'s come from the spectral decomposition of the system's Hamiltonian
   $ H_S = \sum_j \epsilon_j |\epsilon_j\rangle \langle \epsilon_j|$ and $r_j$'s come from the spectral decomposition of the input state $\rho_S(t) = \sum_j r_j \ket{r_j}\bra{r_j}$.
Furthermore, the eigenvalues \( r_j \) and \( \epsilon_j \) are ordered as
\begin{equation}
    r_1 \geq r_2 \geq r_3 \dots \geq r_d, \quad \text{and} \quad \epsilon_1 \leq \epsilon_2 \leq \epsilon_3 \dots \leq \epsilon_d\, .
    \label{eq_spectral_order}
\end{equation}
This ordering ensures that the lowest energy state has the highest population, making the state passive concerning work extraction. For a single-qubit system with \( H_S = \frac{\omega_0}{2} \sigma_z \), the ergotropy is given by~\cite{tiwari2024, andolina_oqs_battery}  
\begin{equation}
   \mathcal{W}[\rho_S(t)] = \frac{\omega_0}{2} \left[ z(t) + \sqrt{x(t)^2 + y(t)^2 + z(t)^2} \right].
\end{equation}
This expression highlights the role of coherence, through \( x(t),\, y(t) \), as well as population imbalance, via \( z(t) \), in determining the extractable work from a single-qubit system. 
To quantify the contribution of (in-)coherence to ergotropy, (in-)coherent ergotropy was introduced~\cite{Francica_CoherentErgotropy}. Coherent ergotropy is directly related to energetic coherence present in the state $\rho_S(t)$, while incoherent ergotropy represents the work that can be extracted from the system without changing its coherence. The incoherent contribution to ergotropy is evaluated as the maximum extractable work from the dephased state, obtained by removing all coherences through the dephasing map. Further, the ergotropy can be written as the sum of its incoherent $(\mathcal{W}_i)$ and coherent parts $(\mathcal{W}_c)$~\cite{Francica_CoherentErgotropy}, i.e., $\mathcal{W} = \mathcal{W}_i +\mathcal{W}_c$, where

\begin{equation}
    \mathcal{W}_i[\rho_S(t)] = \left\{
                                    \begin{array}{ll}
                                        0, & \text{if } z(t) < 0, \\
                                         \omega_0 z(t), & \text{if } z(t) \geq 0.
                                    \end{array}
                                \right.
\end{equation} and
\begin{equation}
    \mathcal{W}_c[\rho_S(t)] = \left\{  
                                       \begin{array}{ll}
                                          \mathcal{W}[\rho_S(t)]  & \text{if } z(t) < 0, \\
                                           \frac{\omega_0}{2} \left[ \sqrt{x(t)^2 + y(t)^2 + z(t)^2} - z(t) \right] & \text{if } z(t) \geq 0.
                                       \end{array}
                                \right.
\end{equation}

Ergotropy is used to characterize the charging and discharging behavior of the quantum battery. The quantum battery is said to have charged when ergotropy increases, and correspondingly, the battery discharges when the ergotropy decreases. The time derivative of ergotropy is the charging power given by
\begin{align}
      \mathcal{P}(t) &= \frac{d \mathcal{W}[\rho_S(t)]}{d t} \nonumber \\
                  &= \frac{\omega_0}{2} \left[\dot z(t) + \frac{x(t) \dot x(t) + y(t) \dot y(t) + z(t) \dot z(t)}{\sqrt{x(t)^2 + y(t)^2 + z(t)^2}}   \right] \nonumber \\
                  &= P(t)\left[1 + \frac{x(t) \dot x(t) + y(t) \dot y(t) + z(t)\dot z(t)}{\dot z(t) \sqrt{x(t)^2 + y(t)^2 + z(t)^2}} \right],
\end{align}
where $P(t)$ is the instantaneous power, Eq.~\eqref{eq_inst_power}. A quantum battery has a positive charging power when it charges, whereas the charging power is negative when the battery discharges. A zero charging power indicates that the battery is neither charging nor discharging. Further, the above equation brings out an interesting relationship between the charging and the instantaneous power, showing that both are connected by a non-zero factor $\left[1 + \frac{x(t) \dot x(t) + y(t) \dot y(t) + z(t)\dot z(t)}{\dot z(t) \sqrt{x(t)^2 + y(t)^2 + z(t)^2}} \right]$.  
\begin{figure}
    \centering
    \includegraphics[width=1\linewidth]{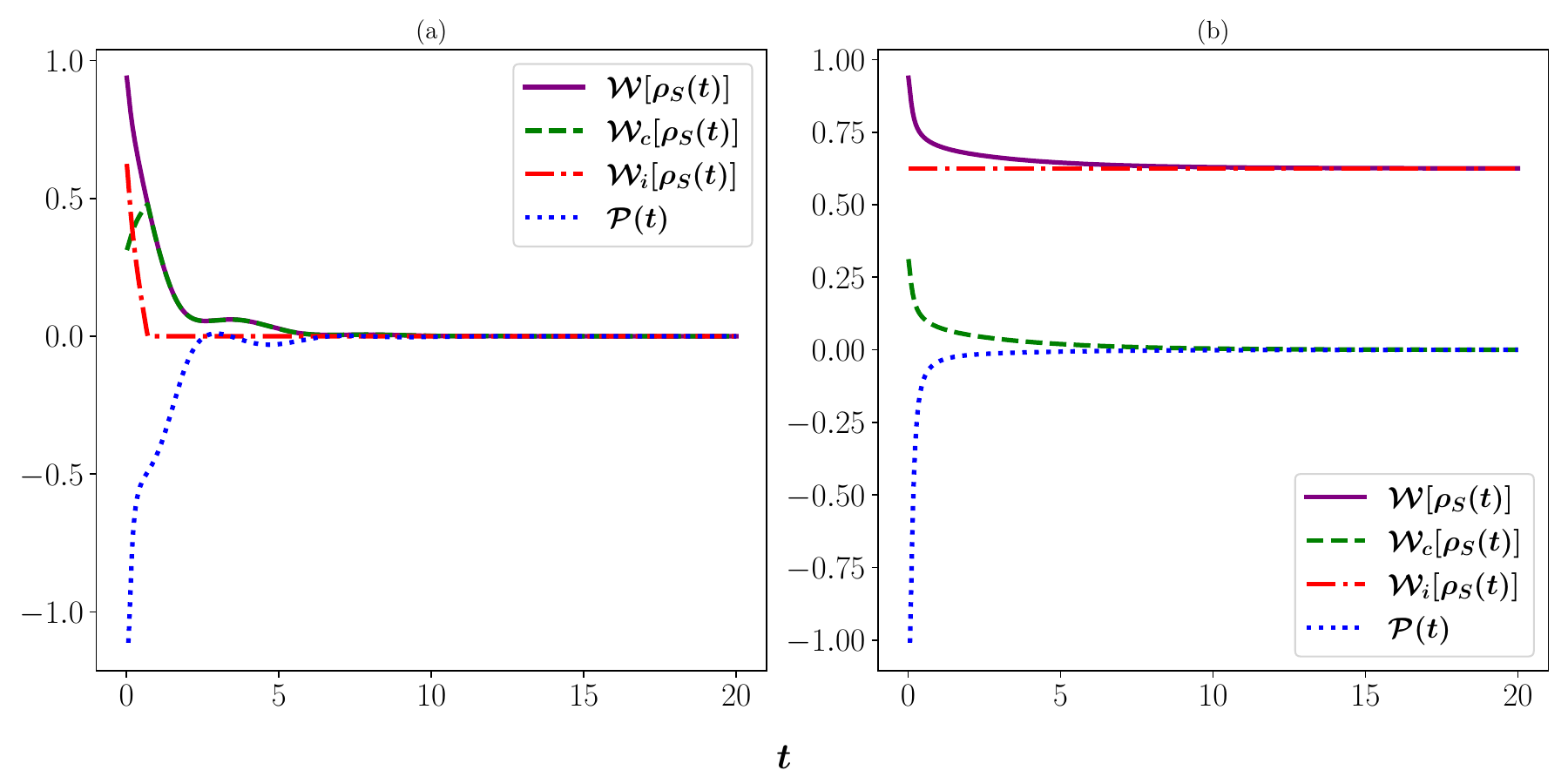}
    \caption{Variation of ergotropy $\mathcal{W}$, (in-)coherent ergotropy $\mathcal{W}_{(i)c}$, and charging power $\mathcal{P}(t)$ with time for both (a) dissipative ($\theta= 0$), and (b) pure dephasing ($\theta = \pi/2$) channels. The parameters are taken to be $\omega_0 = 1.25$, $T = 0.2$, $\Omega = 15$, and $m\gamma = 0.4$.}
    \label{fig_ErgCohINcohPower}
\end{figure}

In Fig.~\ref{fig_ErgCohINcohPower}, we plot ergotropy, its (in-)coherent parts, and the charging power of the quantum battery with time for both dissipative, Fig.~\ref{fig_ErgCohINcohPower}(a), and pure dephasing, Fig.~\ref{fig_ErgCohINcohPower}(b), channels. In the case of pure dephasing evolution, the ergotropy attains a constant, nonzero value after an initial decay. This denotes that, in this case, the battery initially discharges and then stabilizes at a finite value. Further, the contribution to the ergotropy in the long-time limit comes purely from the incoherent ergotropy, as the population (diagonal) terms of the system's density matrix are not impacted during the pure dephasing evolution. The coherent ergotropy dissipates as the system evolves. Further, the charging power is initially negative and reaches zero in the long-time limit when the quantum battery stabilizes. 
\begin{figure*}
    \centering
    \includegraphics[width=0.95\linewidth]{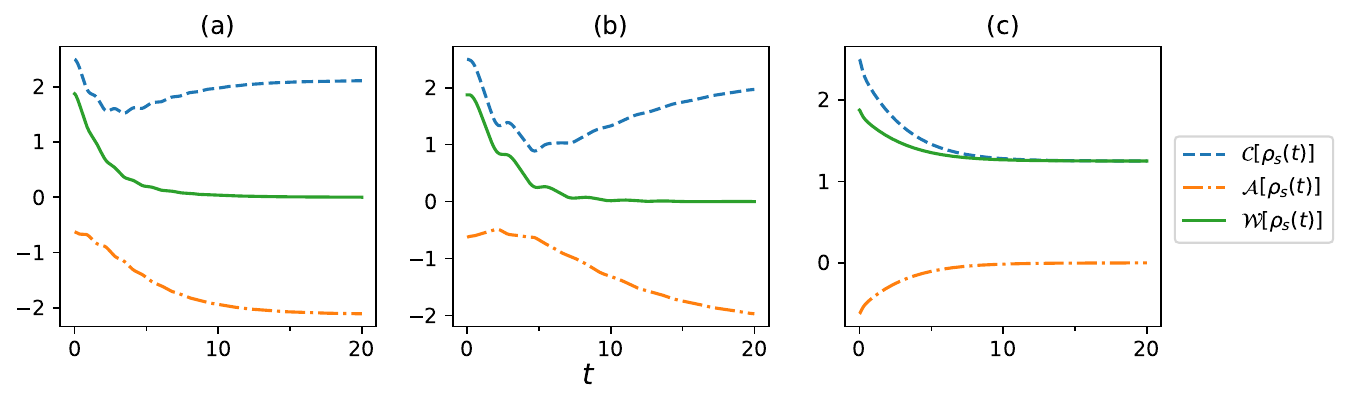}
    \caption{Variation of ergotropy $\mathcal{W}$, anti-Ergotropy $\mathcal{A}$ and battery capacity $\mathcal{C}$ with time for (a) dissipative ($\theta = 0$), (b) mixed ($\theta = \pi/4$), and (c) pure dephasing ($\theta = \pi/2$) evolution. The parameters are taken to be $\omega_0 = 2.5$, $T = 1$, $\Omega = 10$, and $m\gamma = 0.1$.}
    \label{fig_BatteryCapcity}
\end{figure*}
In the case of the dissipative evolution ($\theta = 0$), we observe that the incoherent ergotropy decays sharply as the evolution begins, and the coherent ergotropy becomes equal to the ergotropy of the system. Here, the ergotropy revives after an initial decay due to the non-Markovian evolution of the system. This is reminiscent of the non-Markovian amplitude-damping channel~\cite{tiwari_impact1, Tiwari_impact2}. The charging power is initially negative when the battery discharges, and it becomes positive for a short duration before becoming zero. In the dissipative evolution, the ergotropy decays to zero in the long-time limit, rendering the battery in a passive state.

\subsection{Battery capacity}
A recently developed figure of merit of a quantum battery is its battery capacity~\cite{battery_capacity}. It is a unitarily invariant function of the input state that links the system's work storage capacity to quantum entropies and coherence. It represents the amount of work that a quantum battery can transfer during any thermodynamic cycle using the unitary evolution of the battery. It is the difference between the ergotropy $\mathcal{W}$ (discussed above) and anti-ergotropy $\mathcal{A}$ of the quantum battery, where the anti-ergotropy
\begin{align}
\mathcal{A}[\rho_S (t)] = \text{Tr} \left[\rho_S (t) H_S \right] - \text{Tr} \left[\rho_S^a H_S \right],
\end{align}
is the difference between the energy of the state of the battery and its corresponding active state $\rho_S^a$, given by
\begin{align}
    \rho_S^a = \sum_{j=1}^{d} r_{d+1-j} |\epsilon_j\rangle \langle \epsilon_j|.
\end{align}
Here, $r_j$'s are the eigenvalues of the state $\rho_S(t)$ and $\ket{\epsilon_j}$ are the eigenvectors corresponding to eigenvalues $\epsilon_j$ of the system's Hamiltonian $H_S$. The order of $r_j$ and $\epsilon_j$ are as in Eq.~\eqref{eq_spectral_order}. The active state denotes the highest energy the state can get via unitary evolution. This leads to a bound on the anti-ergotropy $\mathcal{A}\leq 0$. Anti-ergotropy's magnitude denotes how much the quantum battery can be charged. For the state of the quantum battery in Eq.~\eqref{eq_Bloch_vector_form}, we get the following form of anti-ergotropy
\begin{equation}
   \mathcal{A}[\rho(t)] = \frac{\omega_0}{2} \left[ z(t) - \sqrt{x(t)^2 + y(t)^2 + z(t)^2} \right].
\end{equation}
Now, the battery capacity, as the difference between ergotropy and anti-ergotropy, can be written as 
\begin{align}
    \mathcal{C}[\rho_S (t)] = \mathcal{W}[\rho_S (t)] - \mathcal{A}[\rho_S (t)] =   \text{Tr} \left[\rho_S^a H_S \right] - \text{Tr} \left[\rho_S^p H_S \right],
\end{align}
which, for the quantum battery considered here, becomes
\begin{equation}
   \mathcal{C}[\rho(t)] = \omega_0 \left[  \sqrt{x(t)^2 + y(t)^2 + z(t)^2} \right].
\end{equation}
Immediately, from above, we observe that if the battery is in a maximally mixed state, the battery capacity becomes zero. Battery capacity, in general, denotes the energy gap between the maximum possible energy of the battery, calculated using the active state, and the energy of the passive state. 

In the open quantum system considered here, similar to the ergotropy formalism, we feed the state at time $t$ in the evolution as the input state to the battery capacity. From this state, active and passive states are obtained using its spectral decomposition and the spectral decomposition of the system's Hamiltonian. At any time $t$, the maximum work that can be extracted from this open quantum system (quantum battery) unitarily is bounded by ergotropy. The amount to which this battery can be charged via unitary evolution is bounded by anti-ergotropy. The difference between ergotropy and anti-ergotropy provides a bound on battery capacity. To summarize, the battery is, in general, interacting with a bath, and it is losing its charge. During its evolution, if we want to use it, we detach it from the bath and perform unitary evolutions to extract the work (bounded by ergotropy) and charge it (bounded by anti-ergotropy) if needed. 

In Fig.~\ref{fig_BatteryCapcity}, we plot the variation of battery capacity $\mathcal{C}$, anti-ergotropy $\mathcal{A}$ with time for pure dephasing, dissipative, and mixed evolutions of the quantum battery via the master equation~\eqref{gen_spin-boson_master_eq}. Also, we compare them with the evolution of the ergotropy. For the pure dephasing evolution ($\theta = \pi/2$), Fig.~\ref{fig_BatteryCapcity}(c), both the ergotropy and the battery capacity initially decay. However, over an extended period, they converge to an equal, nonzero steady value. Correspondingly, the magnitude of anti-ergotropy, in this case, initially decays but becomes zero at longer times. The behavior of the battery capacity, anti-ergotropy, and ergotropy is almost similar in the case of dissipative ($\theta = 0$) and mixed ($\theta = \pi/4$) evolution. In both, the ergotropy initially decays and becomes zero over time. A slight difference between them is that the revivals in ergotropy for the mixed evolution are more pronounced in the case when $\theta = \pi/4$ as compared to the $\theta = 0$ case. In both cases, we observe that the magnitude of anti-ergotropy increases with time and saturates at longer times. Consequently, the battery capacity, in both cases, first decays, but it increases as time passes, and at longer times, it saturates around two. Interestingly, this value is higher than the saturation value of battery capacity we get in the pure dephasing case for similar parameters. 

Further, to shed light on the storage of the work, we observe the evolution of the battery in the long-time limit. If we want to extract some work from the battery, pure dephasing evolution will be beneficial, as in this case, there will be a finite ergotropy, as can be seen from Fig.~\ref{fig_BatteryCapcity}(c). In contrast, for the dissipative evolution, the ergotropy in the long-time limit will be zero, Figs.~\ref{fig_BatteryCapcity}(a) and (b). However, if we need to recharge the battery, it can be charged to a higher value when the battery is evolved via the dissipative channel, as can be observed from the behavior of anti-ergotropy in Fig.~\ref{fig_BatteryCapcity}.

The information obtained by exploring the dynamic features of the model, such as distinguishability between dissipative and pure dephasing dynamics in the non-Markovianity measure, quantum speed limit, and coherence, has led us to investigate the battery in both these regimes separately. Due to this, distinct advantages of the battery have been obtained in both regimes. It has been outlined that the pure dephasing evolution is better for storing the charge, while dissipative evolution favors recharging. Furthermore, the presence of non-zero coherence enables a positive coherent ergotropy. The non-Markovian evolution has been found to facilitate the recharging of the quantum battery. This has been previously observed for the non-Markovian amplitude damping model~\cite{tiwari_impact1}, where it has been shown that a discharging-charging cycle can be determined by a change in the quantum speed limit time.

\section{Conclusions}\label{conclusions}
In this paper, the dynamic and thermodynamic properties of the spin-boson model in the weak coupling regime were studied. In general, using the Born-Markov and rotating wave approximations, the quintessential dynamics is given by the GKSL master equation. Here, we made use of two master equations. The first one was a weak-coupling spin-boson (WCSB) master equation obtained by relaxing the Markov and rotating wave approximations, and the second was a phase covariant (PC) master equation obtained by applying the rotating wave approximation to the WCSB master equation. The PC master equation has been a topic of interest in recent times. The time-dependent dissipative coefficients for the above master equations were derived using the Drude-Lorentz spectral density. Further, the angle between the direction of the coupling operator in the $xz-$plane and the $x-$axis was utilized to study pure dephasing and dissipative dynamics of the system. First, we studied the dynamics of the model using the WCSB and the PC master equations for both (non-)unital dynamics. The memory effects in the system (non-Markovian evolution) were investigated using the BLP and the RHP measures. The dynamics of the system, when evolved using the WCSB master equation, showed non-Markovian behavior using the BLP measure. Interestingly, however, for the PC master equation, the evolution was BLP Markovian but RHP non-Markovian. In both cases, pure dephasing exhibited Markovian evolution. Further, we investigated the quantum speed limit and the behavior of coherence in the system. Thermal equilibrium and the steady state properties of the system were also examined. It was found that in the steady state, the system attains a thermal state corresponding to the bare system Hamiltonian, consistent with the thermal equilibrium behavior in the weak coupling regime. 

An important aspect of the study was to demonstrate the performance of the spin-boson model as a quantum battery. To characterize the quantum battery, we used quantifiers such as energy, instantaneous power, ergotropy, and its (in-)coherent parts, charging power, anti-ergotropy, and battery capacity. The system was modeled as a quantum battery, and the bosonic bath acted as a charger or dissipator. The impact of the pure dephasing and dissipative evolutions on the battery was clearly brought out. The energy and the power of the battery remained constant in the pure dephasing evolution and decayed in the dissipative regime. In the latter case, the major contribution to ergotropy came from its coherent part (and it became zero in the long-time limit), whereas in the former case, in the long-time limit, ergotropy remained non-zero and equal to the incoherent ergotropy. The magnitude of the anti-ergotropy increased under dissipative evolution, and at longer times, it saturated, whereas for pure dephasing evolution, it gradually became zero. Consequently, the battery capacity was seen to increase in the long-time limit of the dissipative evolution after an initial decay, whereas it became equal to ergotropy after initial decay for the pure dephasing evolution. This study revealed that pure dephasing is better for charge storage in a battery interacting with its environment, as it preserves finite ergotropy over the long time. In contrast, a dissipative evolution boosted the battery's charging capacity in the long-time limit. These insights highlighted the spin-boson quantum battery's critical role in optimizing energy storage and transfer in quantum thermodynamic systems.



\onecolumngrid
\appendix
\section{Calculations involving the factor $\xi(\zeta, t)$}\label{appendix_A}
In Sec.~\ref{spin-boson-model}, we talk about the rates $\gamma_{kj}$, Eq.~\eqref{init_rates}, which involves the factor $\xi(\zeta, t) = \int_0^t ds ~e^{i\zeta s} C(s)$, for $\zeta = \pm\omega_0$ and 0. Here, we present the forms of the real and imaginary parts of the $\xi(\zeta, t)$ for the spectral density given in Eq.~\eqref{drude-Lorentz_spec_dens}. To this end, we use the real and imaginary parts of the bath correlation function $C(t)$, Eqs.~\eqref{re_Ct} and~\eqref{im_Ct}, such that $\alpha_R(\zeta, t) = \int_0^t ds ~e^{i\zeta s}\Re\{C(s)\}$ and $\alpha_I(\zeta, t) = \int_0^t ds ~e^{i\zeta s}\Im\{C(s)\}$.
Now, 
\begin{align}
     \xi(\zeta, t) &=  \alpha_R(\zeta) + i\alpha_I(\zeta),
     \label{app_xi_0_t}
\end{align}

where
\begin{align}
       \alpha_R(\zeta) &= 2m\gamma \Omega^2 T \left( \frac{\Omega + i \zeta}{\Omega (\Omega^2 + \zeta^2)} \left[ 1 - e^{-(\Omega - i \zeta) t} \right] + 2 \sum_{n=1}^{\infty} \frac{1}{\Omega^2 - \nu_{n}^2} \left[ \frac{\Omega (\Omega + i \zeta)}{\Omega^2 + \zeta^2}\left( 1 - e^{-(\Omega - i \zeta )t} \right)  + \frac{\nu_{n}(\nu_{n} + i \zeta)}{\nu_{n}^2 + \zeta^2} \left( e^{-(\nu_n - i \zeta) t} -1\right) \right] \right), \nonumber \\
         \alpha_I(\zeta) &= m \gamma \Omega^2 \frac{\Omega + i \zeta}{\Omega^2 + \zeta^2} \left( e^{-(\Omega - i \zeta) t} -1 \right).
    \label{Xi_omega_0}
\end{align}%
Further, the real and the imaginary parts of $\xi(\zeta, t)$ are given by
\begin{align}
    \Re \left\{\xi(\zeta, t)\right\} &= \Re \left \{ \alpha_R(\zeta) + i\alpha_I(\zeta) \right \},
\end{align}
where
\begin{align}
     \Re \left \{ \alpha_R(\zeta)\right \} &= 2 m \gamma \Omega^2 T \left( \frac{1}{\Omega (\Omega^2 + \zeta ^2)} \left[ \Omega + e^{-\Omega t}(\zeta \sin{\zeta t} - \Omega \cos{\zeta t}) \right]                      \right.\nonumber \\
    &+ \left. 2 \sum_{n = 1}^{\infty} \frac{1}{\Omega^2 - \nu_{n}^2} \left[ \frac{\Omega}{\Omega^2 + \zeta^2}  \left\{  \Omega + e^{-\Omega t}(\zeta \sin{\zeta t} - \Omega \cos{\zeta t})   \right\}  + \frac{\nu_{n}}{\nu_n^2  + \zeta^2} \left\{ e^{-\nu_n t} ( \nu_n \cos{\zeta t} - \zeta \sin{\zeta t}) - \nu_n \right\} \right]    \right), \nonumber \\
    \Re \left \{i \alpha_I(\zeta)\right \} &= \frac{m \gamma \Omega^2 }{\Omega^2 + \zeta^2} \left( e^{-\Omega t} (-\zeta \cos{\zeta t} - \Omega \sin{\zeta t}) + \zeta \right),
    \label{} 
\end{align}
and
\begin{align}
    \Im \left\{\xi(\zeta, t)\right\} &= \Im  \left \{ \alpha_R(\zeta) + i\alpha_I(\zeta) \right \},
\end{align}
where
\begin{align}
    \Im \left \{ \alpha_R(\zeta) \right \} &= 2 m \gamma \Omega^2 T \left(  \frac{1}{\Omega (\Omega^2 + \zeta^2)}  \left[ \zeta - e^{-\Omega t}(\Omega \sin{\zeta t} + \zeta \cos{\zeta t})   \right]   \right.\nonumber \\
    &\quad + \left. 2 \sum_{n=1}^{\infty} \frac{1}{\Omega^2 - \nu_n^2} \left[  \frac{\Omega}{\Omega^2 + \zeta^2} \left\{ \zeta - e^{-\Omega t} (\Omega \sin{\zeta t}  + \zeta \cos{\zeta t}) \right\} + \frac{\nu_n}{\nu_n^2 + \zeta^2} \left\{ e^{-\nu_n t} (\nu_n \sin{\zeta t} + \zeta \cos{\zeta t}) - \zeta \right\} \right]    \right), \nonumber \\
    \Im \left \{ i\alpha_I(\zeta) \right \} &= \frac{m \gamma \Omega^2}{\Omega^2 + \zeta^2} \left( e^{-\Omega t} (\Omega \cos{\zeta t} - \zeta \sin{\zeta t}) - \Omega   \right).
    \label{imagXi}
\end{align}

\section{A discussion on the master equation derived from the spin-boson model in the weak coupling regime}

In this section, we establish the equivalence between the time-convolutionless (TCL) master equations presented in ~\cite{Goan_2010, Goan2011-ke} and in~\cite{Haase2018}. We begin by considering the form of the Hamiltonian provided in~\cite{Goan_2010, Goan2011-ke} (for $\hbar = 1$)
\begin{align}
    H = H_S + H_I + H_R = H_S + \sum_{\lambda}g_\lambda\left(L^\dagger a_\lambda + L a^\dagger_\lambda\right) + \sum \omega_\lambda a^\dagger_\lambda a_\lambda,
    \label{app_Hamiltonian}
\end{align}
where $H_S$, $H_R$, and $H_I$ are the system, bath, and system-bath interaction Hamiltonians, respectively. The system Hamiltonian is given by $H_S = \frac{\omega_0}{2}\sigma_z$. 
The corresponding second-order TCL non-Markovian master equation for the reduced density matrix $\rho_S(t)$ was derived to be~\cite{Goan_2010, Goan2011-ke}
\begin{equation}
    \frac{d\rho_S(t)}{dt} = -i \left[ H_S, \rho_S(t)  \right]  - \int_{0}^{t} d\tau \left\{ \alpha(t -\tau) \left( L^{\dagger}\widetilde{L}(\tau - t) \rho_S(t) -  \widetilde{L}(\tau - t) \rho_S(t) L^{\dagger}  \right)  + \beta(t-\tau) \left( L \widetilde{L}^{\dagger}(\tau - t)\rho_S(t) -  \widetilde{L}^{\dagger}(\tau - t)\rho_S(t) L  \right)  + h.c. \right\},
    \label{nMEq}
\end{equation}
where $h.c.$ indicates Hermitian conjugate of previous terms, and $ \widetilde{L}(t)$ is the system operator in the interaction picture with respect to system Hamiltonian $H_S$ 
\begin{equation}
    \widetilde{L}(t) = e^{i H_S t}L e^{-i H_S t}.
\end{equation}
The correlation functions $\alpha(\tau -s)$ and $\beta(\tau -s)$, see Eq.~\eqref{C_S} and Ref.~\cite{Goan_2010, Goan2011-ke}, are defined as
\begin{align}
    \alpha(t - \tau) &= \sum_{\lambda} (\overline{n}_{\lambda} + 1) |g_{\lambda}|^{2} e^{-i\omega_{\lambda}(t - \tau)}, \nonumber \\
    \beta(t - \tau) &= \sum_{\lambda} \overline{n}_{\lambda} |g_{\lambda}|^{2} e^{i\omega_{\lambda}(t - \tau)}, \nonumber \\
    \alpha_{eff}(\tau -s) &= \alpha(t - \tau) + \beta(t - \tau) \nonumber\\
                          &= \sum_{\lambda} |g_{\lambda}|^{2} \left[ \left( 1 + 2\overline{n}_{\lambda} \right) \cos{\omega_{\lambda}(t - \tau)} -i \sin{\omega_{\lambda}(t - \tau)}\right],
    \label{C_F}
\end{align}
Here, $g_{\lambda}$ and $\omega_{\lambda}$ are the coupling strength and the frequency of the $\lambda$th environment oscillator, respectively. The thermal mean occupation number of bosons or harmonic oscillators is $\overline{n}_{\lambda} = \left[ \exp\left({\frac{ \omega_{\lambda}}{k_{B}T}}\right) - 1 \right]^{-1}$. Further, using the bath spectral density of the form $J(\omega) = \sum_\lambda |g_\lambda|^2\delta(\omega - \omega_\lambda)$, the function $\alpha_{eff}(t - \tau)$ can be written as 
\begin{align}
    \alpha_{eff}(t - \tau) = \int_0^\infty d\omega J(\omega)\left\{\coth\left(\frac{\omega}{2T}\right)\cos\left[\omega(t - \tau)\right] - i\sin\left[\omega(t - \tau)\right]\right\}.
\end{align}
By comparing the above correlation function with the correlation function discussed in Eq.~\eqref{C_S}, it can be seen that $\alpha_{eff}(t - \tau) \equiv C(t - \tau)$. 

Now, we demonstrate the equivalence between the WCSB master equation [Eq.~\eqref{gen_spin-boson_master_eq}] and Eq.~\eqref{nMEq} by deriving the former from the latter, provided the Hamiltonians governing their dynamics are equivalent. Comparing the Hamiltonian in Eq.~\eqref{app_Hamiltonian} with the Hamiltonian in Eq.~\eqref{total_hamiltonian} reveals that they match when the operator $L$ in Eq.~\eqref{app_Hamiltonian} is chosen as $L = \cos\left(\theta\right)\frac{\sigma_x}{2} + \sin\left(\theta\right)\frac{\sigma_z}{2}$. It implies $L = L^\dagger$ and 
\begin{align}
    \widetilde L(t - \tau) = \frac{\cos\theta }{2}\left(e^{i\omega_0(t-\tau)}\sigma_+ + e^{-i\omega_0(t-\tau)}\sigma_-\right) + \frac{\sin \theta}{2}\sigma_z\, , && \text{and} && \widetilde L^\dagger(t - \tau) = \widetilde L(t - \tau).
\end{align}
Further, using the properties of $L$ and $\widetilde L(t-\tau)$, we can simplify Eq.~\eqref{nMEq} to be
\begin{align}
    \frac{d\rho_S(t)}{dt} = -i \left[ H_S, \rho_S(t)  \right]  - \int_0^td\tau\left\{\alpha_{eff}(t - \tau)\left[L \widetilde L(\tau - t)\rho_S(t) - \widetilde L(\tau - t)\rho_S(t)L\right] + \alpha_{eff}^*(t-\tau)\left[\rho_S(t)\widetilde L(\tau - t) L - L\rho_S(t)\widetilde L(\tau - t)\right]\right\}.
    \label{app_NMeq_2}
\end{align}
We now make the following substitutions in the above integration: $ t-\tau = s$, implying $d\tau = -ds$, and thereby making $\alpha_{eff}(t-\tau) = \alpha_{eff}(s) = C(s)$. Further, $\widetilde L(\tau - t)$ in the above equation becomes 
\begin{align}
    \widetilde L(\tau - t) = \widetilde L(-s) = \frac{1}{2}\left\{\cos(\theta)\left(e^{-i\omega_0s}\sigma_+ + e^{i\omega_0s}\sigma_-\right) + \sin(\theta)\sigma_z\right\}. 
    \label{app_eq_Ls}
\end{align}
To this end, Eq.~\eqref{app_NMeq_2} can be simplified to be 
\begin{align}
    \frac{d\rho_S(t)}{dt} = -i \left[ H_S, \rho_S(t)  \right]  + \int_0^tds\left\{C(s)\left[L \widetilde L(-s)\rho_S(t) - \widetilde L(-s)\rho_S(t)L\right] + C^*(s)\left[\rho_S(t)\widetilde L(-s) L - L\rho_S(t)\widetilde L(-s)\right]\right\}.
\end{align}
Upon substituting $L(-s)$ from Eq.~\eqref{app_eq_Ls} and $L = \frac{1}{2}\left\{\cos(\theta)(\sigma_+ + \sigma_-) + \sin(\theta) \sigma_z\right\}$ in the above equation, we get the form of a master equation equivalent to Eq.~\eqref{gen_spin-boson_master_eq}. 
\section{Thermodynamic features of the system evolved via the PC master equation}
\begin{figure}
    \centering
    \includegraphics[width=0.95\linewidth]{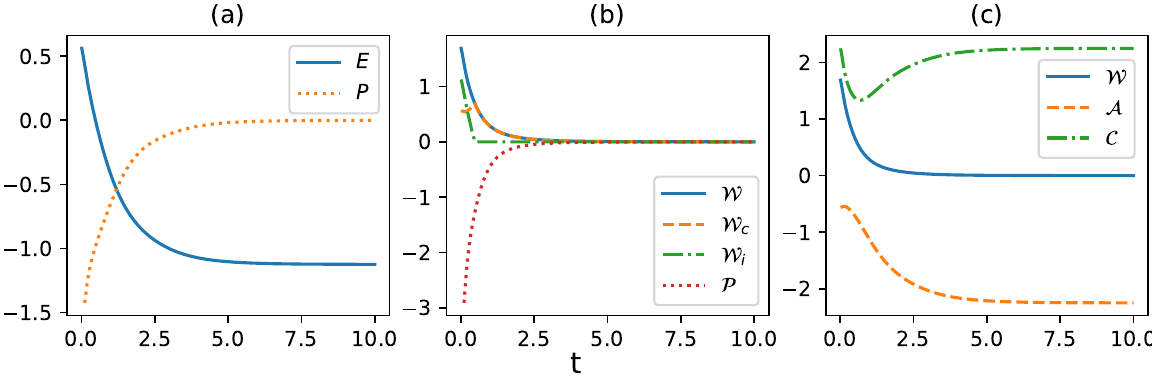}
    \caption{Variation of (a) energy, instantaneous power, (b) ergotropy, (in-)coherent ergotropy, charging power, and (c) battery capacity, (anti)-ergotropy with time for dissipative case ($\theta=0$) in the PC dynamics, Eq.~\eqref{PC_dynamics}. The parameters are taken to be $ T = 0.2$, $\Omega = 15$, $m \gamma = 0.4$ and $\omega_0 = 2.25$.}
    \label{fig_PCCappendixv2}
\end{figure}
Here, we present the quantum thermodynamic properties of the spin-boson model (the system is modeled as a quantum battery) for the evolution of the system being governed by the PC master equation, Eq.~\eqref{PC_dynamics}. The initial state of the system is taken to be $\frac{\sqrt{3}}{2}\ket{0} + \frac{1}{2}\ket{1}$. In Fig.~\ref{fig_PCCappendixv2}, we present the variation of the quantum battery characteristic quantifiers with time. From Fig.~\ref{fig_PCCappendixv2}(a), we find that the energy of the quantum battery decreases monotonically until it reaches the ground state $\frac{-\omega_0}{2}$, where it saturates. The corresponding instantaneous power is negative and later becomes zero. The variation of the ergotropy, its (in-)coherent parts, and the charging power with time is shown in Fig.~\ref{fig_PCCappendixv2}(b). It can be observed that the incoherent ergotropy decays sharply and becomes zero. As a consequence, the coherent ergotropy becomes equal to the ergotropy of the system. The ergotropy of the system decreases monotonically and becomes zero after some time. A comparison between the ergotropy, anti-ergotropy, and battery capacity is illustrated in Fig.~\ref{fig_PCCappendixv2}(c). The magnitude of the anti-ergotropy is seen to increase with time, and the ergotropy gradually becomes zero. Due to this, the battery capacity initially decays but later increases and saturates at the initial value. 
\twocolumngrid

\bibliography{reference}
\bibliographystyle{apsrev}

\end{document}